\documentclass[prb,twocolumn,superscriptaddress,showpacs]{revtex4-1}

\usepackage{amsmath}
\usepackage{amssymb}
\usepackage{graphicx}
\usepackage{hyperref}
\usepackage{color}
\usepackage{simplewick}

\DeclareMathOperator{\e}{e}

\newcommand{\up}{\uparrow}
\newcommand{\dw}{\downarrow}
\newcommand{\bB}{\mathbf{B}}
\newcommand{\bS}{\mathbf{S}}
\newcommand{\Ket}[1]{|#1\rangle}
\newcommand{\Bra}[1]{\langle#1|}
\newcommand{\Bracket}[2]{\langle#1|#2\rangle}
\newcommand{\mean}[1]{\langle#1\rangle}

\begin{document}

\title{Spin filtering and entanglement detection due to spin-orbit interaction in carbon nanotube cross-junctions}

\newcommand{\sns}{NEST, Scuola Normale Superiore, and Istituto Nanoscienze-CNR, I-56126 Pisa, Italy}
\newcommand{\madrid}{Departamento de F\'{\i}sica Te\'{o}rica de la Materia Condensada,
Condensed Matter Physics Center (IFIMAC), and Instituto Nicol\'{a}s Cabrera,
Universidad Aut\'{o}noma de Madrid, E-28049 Madrid, Spain}
\newcommand{\braunschweig}{Institute for Mathematical Physics, Technical University Braunschweig,
Mendelssohnstr.~3, D-38106 Braunschweig, Germany}

\author{Francesco Mazza}
\affiliation{\madrid} 
\affiliation{\sns}

\author{Bernd Braunecker}
\affiliation{\madrid}

\author{Patrik Recher}
\affiliation{\braunschweig}

\author{Alfredo Levy Yeyati}
\affiliation{\madrid}



\begin{abstract}
We demonstrate that, due to their spin-orbit interaction, carbon nanotube cross-junctions 
have attractive spin projective properties
for transport. First, we show that the junction can be used 
as a versatile spin filter as a function of a backgate and a static external magnetic field.
Switching between opposite spin filter directions can be achieved by small changes of 
the backgate potential, and a full polarization is generically obtained in an energy range
close to the Dirac points. Second, we discuss how the spin filtering properties
affect the noise correlators of entangled electron pairs, which allows us
to obtain signatures of the type of entanglement that are different from the signatures
in conventional semiconductor cross-junctions.
\end{abstract}


\maketitle


\section{Introduction}

Over the last two decades, carbon nanotubes (CNTs) have developed into a mature
material that can be produced at high purity,\cite{tans:1997,cobden:2002,liang:2002,minot:2004,jarillo-herrero:2004,jarillo-herrero:2005,graeber:2006}
and important steps toward mass production have been taken.\cite{vijayaraghavan:2012,park:2012}
This makes CNTs an attractive platform for a future implementation of quantum information processing or spintronics. 
In this context, CNTs have already been proven functional for producing correlated electron pairs 
in a double quantum dot Cooper pair splitter setup.\cite{hermann:2010,schindele:2012} 
Such Cooper pair splitters\cite{recher:2001} are first implementations of a source of entangled electron
pairs on demand, complementing similar implementations in semiconductors,\cite{hofstetter:2009,hofstetter:2011,das:2012} 
and they are related closely to proposals for generating entangled pairs in systems 
with forked geometry.\cite{torres:1999,deutscher:2000,lesovik:2001,recher:2002,bena:2002,recher:2003,hu:2004,samuelsson:2004,sauret:2005,bayandin:2006}

To actively control the electron spin, spin-orbit interaction (SOI) effects 
have found much attention in recent years, since they allow an all-electric local control of the 
electron spin. While in semiconductors the SOI typically causes the 
spin to precess during transport, in CNTs, which are hollow cylinders different from
filled quantum wires, the SOI has a different impact. 
Rather than causing spin rotations
it lifts the energy degeneracy of opposite spins and leads to distinct, fully spin polarized bands with the polarization 
directions parallel to the rotational symmetry axis,\cite{ando:2000,chico:2004,huertas-hernando:2006,izumida:2009,jeong:2009,chico:2009,klinovaja:2011a,klinovaja:2011b,delvalle:2011,chico:2012} 
an effect whose consequences were to date investigated theoretically mainly in 
quantum dot setups.\cite{bulaev:2008,weiss:2010,burset:2011,cottet:2012a,cottet:2012b}
Recently the possibility of using gates to control the spin filtering properties due to SOI in CNT quantum dots has been also exploited.\cite{braunecker:2013}
While the band splitting properties of the SOI have been confirmed
by experiments,\cite{kuemmeth:2008,jespersen:2011,steele:2013}
the spin-projective properties still require experimental testing.

\begin{figure}
   \includegraphics[width=\columnwidth]{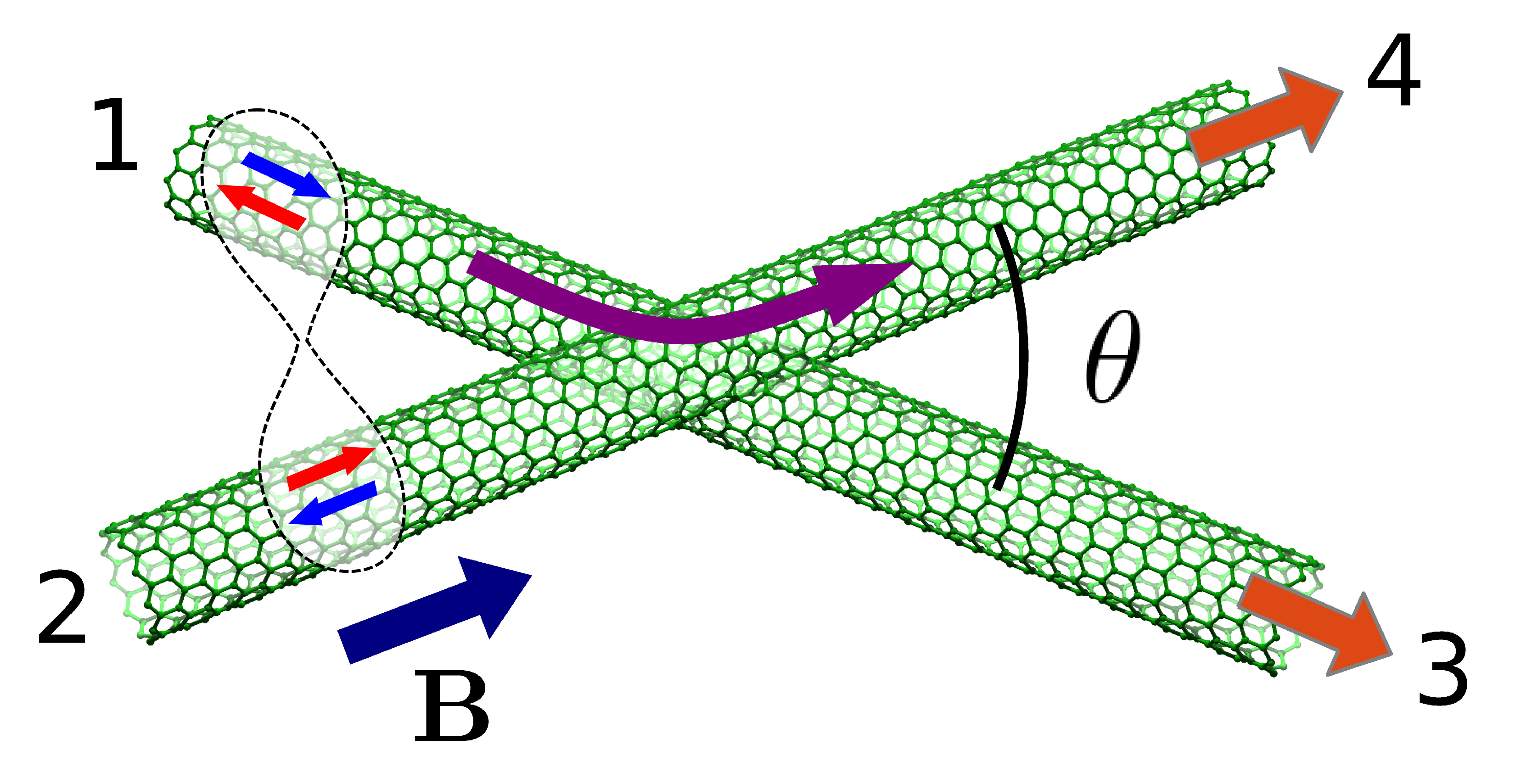}
   \caption{\label{fig:beamsplitter}
   (Color online)
   Two CNTs forming a cross-junction, a four terminal system labeled 
   by the leads $i=1,2,3,4$ with a small contact area that allows electron tunneling between 
   the CNTs, and an angle $\theta$ between the CNT axes. 
   Any injected particle can at the junction either remain in the same CNT or tunnel into 
   the other CNT, illustrated by the bent arrow for the tunneling process $i=1 \to i=4$.
   We assume that particles are injected into leads $i=1,2$ and their current is measured at 
   the outgoing leads $i=3,4$ (orange arrows). A magnetic field $\bB$ is applied in the plane
   spanned by the two CNTs, and we mostly consider the case of $\bB$ parallel to the nanotube
   with leads $i=2,4$.
   This setup is explored for the combined effects of SOI and $\bB$ on
   spin-filtered transport and signatures of entanglement from electron pairs, such as the
   incoming spin-entangled state indicated by the hourglass shape.}
 \end{figure}

In this paper, we show that single-wall CNT cross-junctions as shown in Fig. \ref{fig:beamsplitter}
are attractive candidates for further experimental progress for both spin-resolved
transport and the detection of entanglement signatures.
While such cross-junctions have already attracted much interest 
experimentally\cite{postma:2000,fuhrer:2000,kim:2001,janssen:2002,yoneya:2002,park:2003,gao:2004}
and theoretically,\cite{yoon:2001,nakanishi:2001,komnik:2001,dag:2004,margulis:2007,havu:2011}
the novel features due to SOI have never been investigated before.
Here we take the SOI fully into account, and consider setups with 
weak (usually sub-tesla) magnetic fields 
$\bB$ in the plane spanned by the two CNTs, often with $\bB$ parallel to one of the CNTs.

We first demonstrate that close to the Dirac points the cross-junction operates as 
an efficient spin filter.
Full spin polarization for an outgoing current is obtained
if only one SOI-split subband contributes to the outgoing transport. This polarization can 
be reversed to opposite, yet not perfect polarization upon small changes of the gate potential
such that further subbands become active for the transport. At fixed gate potential, however,
a perfect reversal of the polarization in the lowest subband can be achieved by letting
$\bB \to -\bB$. 

In the second part of this paper, we investigate the transport properties of 
entangled electron pairs passing through the cross-junction, as indicated by the 
hourglass shaped state in Fig. \ref{fig:beamsplitter}. We investigate the current noise
correlators for signatures of the entanglement, notably for a bunching or antibunching behavior 
arising from the injection of singlet or triplet states. This investigation is an extension
of previous work on semiconductor cross-junctions,\cite{burkard:2000,samuelsson:2004} in which also SOI
effects\cite{egues:2002,egues:2005,sanjose:2006} were 
investigated. Due to the different band structure of CNTs and SOI effects, our results
are	quite different, and a comparison will be made accordingly below.

We will consider only cross-junctions of weakly coupled CNTs, allowing us to 
connect the scattering theory of the cross-junction directly to a microscopic Hamiltonian.
Hence we do not treat high-efficiency (50-50) beam splitters such as required for 
proposed noise-measurement based proofs of entanglement through, for instance, 
Bell inequalities,\cite{kawabata:2001,chtchelkatchev:2002,samuelsson:2003}
entanglement witnesses,\cite{faoro:2007}
or quantum state tomography.\cite{samuelsson:2006}
It should also be noted that a proof of entanglement does not necessarily need
a beam splitter setup. If spin filters, for instance, as provided by the SOI-split CNT bands,
are placed close to the source of entangled electron pairs, the 
current amplitude of the outgoing pairs is modulated by the nonlocal filter settings.
Hence entanglement information can be obtained already from measuring 
currents only.\cite{braunecker:2013}

The paper is organized as follows, in Sec. \ref{sec:CNT} we introduce the model for the CNT
including the SOI effects, and we provide the scattering theory description of the cross-junction.
We then analyze in Sec. \ref{sec:cond} the normal state cross-conductance and the 
spin filtering properties. 
In Secs. \ref{sec:inj} to \ref{sec:non_adj} we discuss the noise properties of injected
entangled pairs. Section \ref{sec:inj} contains an analysis of different injection scenarios,
Sec. \ref{sec:noise} the proper analysis of the current noise correlators, 
Sec. \ref{sec:B} the dependence of the noise on varying the external magnetic field,
and Sec. \ref{sec:non_adj} a discussion of the noise properties under non-ideal 
particle injections. We conclude in Sec. \ref{sec:conclusions}. Two appendices
\ref{app:correlators} and \ref{app:identity} contain some details of the calculations.


\section{CNT cross-junctions}
\label{sec:CNT}

\subsection{CNT low-energy Hamiltonian}

CNTs can be considered as graphene sheets that are rolled into a cylinder.\cite{saito:1998}
They inherit from graphene the low-energy band structure in the form of two Dirac valleys
centered at momenta commonly denoted as $K$ and $K'$.
Since momenta are quantized in the transverse (circular) direction, the resulting
CNT low-energy band structure consists of cuts through the Dirac cones, which form subbands
labeled by the quantized transverse momenta $k_\perp$ and described by the
single-particle Hamiltonian
\begin{equation} \label{eq:H0}
	H_0 = \hbar v_F (k_\perp \sigma_1 + k \tau \sigma_2),
\end{equation}
where $\hbar$ is Planck's constant, $v_F \approx 0.9 \times 10^6$ m/s
is the Fermi velocity, $k$ are the longitudinal momenta along the CNT axis,
$\tau = +,- = K,K'$ labels the Dirac valleys, and $\sigma_{1,2}$ are Pauli matrices
referring to the $A,B$ sublattice components of the wave functions.
If the indices $(N_1,N_2)$ denote the chirality of the CNT, i.e.,
how the graphene sheet is rolled together,
we have $k_\perp = [n- (N_1-N_2 \mod 3)/3]/R$,
with the integer $n$ being the subband index, $R = a \sqrt{N_1^2+ N_2^2+ N_1 N_2}$ the
CNT radius, and $a = 2.46$ \AA\ the unit cell length.

The SOI and the hybridization of orbitals induced by the curvature of the surface
of the CNT lead to additions to the Hamiltonian $H_0$
given by,\cite{izumida:2009,jeong:2009,klinovaja:2011a,klinovaja:2011b}
\begin{align}
\label{eq:H_cv}
	H_{cv} &= \hbar v_F ( \Delta k_\perp^{cv} \sigma_1 + \tau \Delta k^{cv}_z \sigma_2 ),
\\
\label{eq:H_SOI}
	H_{SOI} &=  (\alpha \sigma_1 + \tau \beta ) S^z,
\end{align}
with $S^z$ the Pauli matrix for the spin operator parallel to the CNT axis
(all used Pauli matrices are normalized to eigenvalues $\pm 1$),
$\Delta k_{\perp,z}^{cv}$ the curvature induced momentum shifts,
and $\alpha, \beta$ the SOI interaction strengths. The values of these parameters
depend on the precise details of the elementary overlap integrals between the carbon orbitals
and are subject to considerable uncertainty.\cite{serrano:2000,min:2006,guinea:2010}
From a rather conservative estimate of the resulting SOI strength
one obtains\cite{klinovaja:2011a,klinovaja:2011b}
$\hbar v_F \Delta k_\perp^{cv} = -\tau \cos(3\eta) 5.4 \text{meV}/ (R \text{[nm]})^2$,
$\hbar v_F \Delta k_z^{cv}      =  \tau \sin(3\eta) 5.4 \text{meV}/ (R \text{[nm]})^2$,
$\alpha = -0.08 \text{meV}/R \text{[nm]}$, and
$\beta  = -\cos(3\eta) 0.31 \text{meV}/R \text{[nm]}$,
for $\eta$ the chiral angle defined by $\tan\eta = \sqrt{3} N_2/(2N_1+N_2)$.

The application of an external magnetic field $\bB = (B_x,B_y,B_z)$, with $B_z$ parallel
to the CNT axis, leads to the further terms
\begin{equation}
	H_B = \mu_B g \bB \cdot \bS + |e| v_F R B_z \sigma_1/2,
\end{equation}
incorporating the Zeeman effect and the Aharonov-Bohm flux of the magnetic field through the
CNT cross section.
Here $\bS = (S^x,S^y,S^z)$ is the vector of spin Pauli matrices, $\mu_B$ the Bohr magneton,
$g=2$ the Land\'{e} {\it g}-factor, and $e$ the electron charge. We shall consider only weak fields
not exceeding one or a few tesla, allowing us to neglect further
orbital terms that would lead to the formation of Landau levels.

\begin{figure}
   \includegraphics[width=\columnwidth]{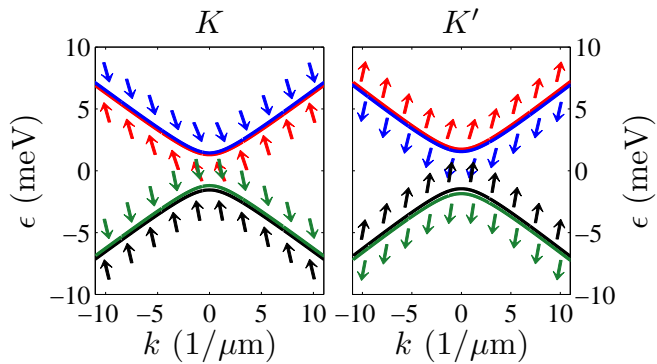}
   \caption{\label{fig:bands}
   (Color online)
   Lowest CNT bands close to the $K$ and $K'$ points for a CNT with chirality (30,12)
   in the presence of a magnetic field $|\bB|=0.6$ T at the angle $\theta = \pi/3$ with respect
   to the CNT axis. The combined effects of CNT curvature, SOI, and magnetic field
   fully gap the nominally metallic CNT and lead to entirely non-degenerate, fully spin 
   polarized bands. The arrows next to the curves indicate the spin polarizations of the 
   bands as a function of momentum $k$, in the plane spanned by $\bB$ and the CNT axis
   (with the axis direction upwards in the plot). At large energies, the polarizations become
   energy independent and follow the effective, valley dependent Zeeman fields 
   $\bB^\text{eff}_\tau = \tau \bB_{SOI} + \bB$.
   The shown SOI induced features are generic for all chiralities, except for armchair
   CNTs, including semiconducting CNTs, with the main difference being different SOI
   interaction strengths.
   }
\end{figure}

Figure \ref{fig:bands} shows a typical spectrum resulting from the combination of SOI and the
external magnetic field. The SOI spin-splits the bands and causes a spin $S^z$ polarization
along the CNT axis. Since the SOI maintains time-reversal symmetry, the bands in the
$K$ and $K'$ valleys remain degenerate but carry opposite spin polarizations.
With the external magnetic field, the time-reversal symmetry is broken, and the
effective Zeeman fields acting in both valleys are different. While for any momentum
$k$ each band remains fully spin polarized, the polarization directions are no longer
opposite in the two valleys and the spins tend to align with the transversal component
of the external field, as shown by the arrows in Fig. \ref{fig:bands}.
We denote the corresponding $(k,\tau)$-dependent spin eigenvalues by $\nu = \pm$.

Away from the Dirac points, the polarizations become $k$ independent and the
effective Zeeman field becomes $\bB^\text{eff}_{\tau,n} = \tau \bB_{SOI,n} + \bB$
with $n$ the band index in each valley and $\bB_{SOI,n}$ the result of the SOI in band $n$.
In addition, the orbital effect of $\bB$ shifts the energy levels such that
both valleys also become energetically nondegenerate. In the low-energy regime close
to the Dirac points, therefore, any transport is strongly subjected to the spin and valley
filtering properties of the bands, together with a strongly enhanced density of states
due to the curved band bottoms. Finally it should be noted that for $\bB$ parallel to the
CNT axis, $S^z$ remains the good spin quantum number for all $k$, and
$\nu$ is identified with the spin projection along the CNT axis.

\subsection{Cross-junction}

The cross-junctions considered in this paper act as scatterers 
transferring incoming electrons
between the two nanotubes and between different bands within
each nanotube. We describe these processes within the scattering matrix
formalism,\cite{buttiker:1992} for which the cross-junction
forms a four terminal system with two incoming $i=1,2$ and two outgoing
leads $i=3,4$, as shown in Fig.~\ref{fig:beamsplitter}.
The scattering states are labeled by the further valley and spin
quantum numbers $\tau,\nu$
and the energies $\epsilon$, and are represented by the states
$\Ket{i,\tau,\nu,\epsilon}$ and the fermion operators $a_{i\tau\nu}^\dagger(\epsilon)$.

We denote by $s_{(i,\tau,\nu),(i',\tau',\nu')}(\epsilon)$ the
scattering matrix for the elastic process
$\Ket{i',\tau',\nu',\epsilon} \to \Ket{i,\tau,\nu,\epsilon}$.
We do not consider inelastic processes because the coherence length is usually larger
than the typical dimension of the junction.

Two scenarios for the scattering matrix will be considered. 
First, we neglect backscattering between the incoming leads $i=1,2$
and the outgoing leads $i=3,4$.\cite{burkard:2000}
In a basis for the leads $i=1,2,3,4$, we then have
\begin{equation} \label{eq:s}
	s =
	\begin{pmatrix}
		0 & 0 & t_{1,3} & r_{1,4} \\
		0 & 0 & r_{2,3} & t_{2,4} \\
		t_{3,1} & r_{3,2} & 0 & 0\\
		r_{4,1} & t_{4,2} & 0 & 0
	\end{pmatrix},
\end{equation}
where $r_{i,i'}$ and $t_{i,i'}$ are matrices in $\tau,\nu,\epsilon$.
This description is appropriate for a larger contact area, and small $\theta$
in which the momentum of the incoming wave packets is approximately preserved.

Second, we focus on a contact area between the two CNTs with a linear extension
much smaller than the size of typical incoming wave packets, which 
allows us to treat the tunnel junction as an elastic point contact scatterer.
Such a situation is typically obtained when one CNT
falls over another CNT.~\cite{fuhrer:2000, park:2003}
Usually then the tunneling coupling is weak, allowing us to retain only first order
tunneling processes and exclude backscattering into the same lead. 
Scattering between leads $1 \leftrightarrow 2$ and $3 \leftrightarrow 4$,
however, must now be taken into account. The scattering matrix becomes
\begin{equation} \label{eq:sp}
	s =
	\begin{pmatrix}
		0 & r_{1,2} & t_{1,3} & r_{1,4} \\
		r_{2,1} & 0 & r_{2,3} & t_{2,4} \\
		t_{3,1} & r_{3,2} & 0 & r_{3,4}\\
		r_{4,1} & t_{4,2} & r_{4,3} & 0
	\end{pmatrix}.
\end{equation}
The consequences of both scenarios are discussed in Sec. \ref{sec:noise}.

With the assumption of a weak tunneling amplitude, we proceed to 
make a Born approximation to link the scattering matrix 
to the microscopic tunneling Hamiltonian $H_t$.
In this approximation, the tunneling between the two CNTs, the scattering
between the leads 
$1\leftrightarrow 4$, $2 \leftrightarrow 3$, $1 \leftrightarrow 2$, and $3 \leftrightarrow 4$ 
is expressed by
\begin{equation}
	r_{(i,\tau,\nu),(i',\tau',\nu')}(\epsilon)
	=
	\Bra{i,\tau,\nu,\epsilon} H_t \Ket{i',\tau',\nu',\epsilon},
\end{equation}
while the transmission within the same nanotube, $1 \leftrightarrow 3$ and $2 \leftrightarrow 4$,
remains unperturbed
\begin{equation}
	t_{(i,\tau,\nu),(i',\tau',\nu')}(\epsilon)
	=
	i \delta_{\tau,\tau'} \delta_{\nu,\nu'}.
\end{equation}
Note that if we choose these matrix elements to be purely imaginary, we can choose real
$r_{i,i'}$ below and maintain an approximate unitarity of the 
scattering matrix in the Born approximation.
The tunneling can be described by a tight binding Hamiltonian of the form\cite{nakanishi:2001}
\begin{align}
	H_t
	&=
	\sum_{\substack{\tau, \tau', \sigma, \sigma'\\s,x_n,x_{n'}}}
	\lambda_{\tau,\sigma; \tau',\sigma'} (x_n,x_{n'})
	\tilde{a}^\dag_{2, \tau, \sigma, s} (x_n) \tilde{a}_{1,\tau',\sigma',s}(x_{n'})
\nonumber \\
	&+ \text{H.c.},
\label{eq:H_t}
\end{align}
where $x_n,x_{n'}$ mark the unit cell positions of the two CNTs at the contact area,
and $s=\up,\dw$ denotes the spin projections in a global spin basis.
The tunneling amplitudes $\lambda_{\tau,\sigma;\tau',\sigma'}(x_n,x_{n'})$
preserve the spin but may be sublattice and valley dependent. The operators $\tilde{a}_{i,\tau,\sigma,s}^\dagger$
are the microscopic electron operators, related to the scattering states by
the transformation
\begin{equation}
	a^\dag_{i, \tau, \nu}(\epsilon)
	= \sum_{x_n,\sigma, s}
	g_{i,\tau,\nu,\epsilon}^{\sigma, s}(x_n)
	\tilde{a}^\dag_{i, \tau, \sigma, s} (x_n),
 \end{equation}
with the transformation matrix $g$. It should be noted that while the valley $\tau$ is preserved,
the sublattice, position, and global spin coordinates are summed out.
Through the summation of the latter, together with the fact that $H_t$ is spin preserving,
the scattering matrix elements $r_{(i,\tau,\nu),(i',\tau',\nu')}$ are proportional
to the spin overlap integral $\Bracket{\nu}{\nu'}$ between the states $\nu$ and $\nu'$ of the two CNTs.
If $\bS_{i,\tau,\nu,\epsilon} = \Bra{i,\tau,\nu,\epsilon} \bS \Ket{i,\tau,\nu,\epsilon}$
is the spin polarization vector of band $(i,\tau,\nu)$ at energy $\epsilon$, this spin
overlap integral allows us to express the scattering matrix elements for the tunneling processes
as
\begin{align}
	&|r_{(i,\tau,\nu),(i',\tau',\nu')}(\epsilon)|^2
\nonumber\\
	&= \Gamma \, \rho_{i,\tau,\nu,\epsilon} \, \rho_{i',\tau',\nu',\epsilon}
	\,
	(1 + \bS_{i,\tau,\nu,\epsilon} \cdot \bS_{i',\tau',\nu',\epsilon}),
\label{eq:r2}
\end{align}
with $\Gamma$ the effective tunnel rate obtained from summing out the $\lambda_{\tau,\sigma;\tau',\sigma'}(x_n,x_{n'})$
in Eq. \eqref{eq:H_t}, and $\rho_{i,\tau,\nu,\epsilon}$ the
density of states in band $(i,\tau,\nu)$. It should be noted that close to a band
bottom, at which one of the involved densities of states diverges, this perturbative formula is
no longer accurate, and the singular behavior is truncated by higher order processes.
However, since the bands are spin projective, the proportionality to
$(1 + \bS_{i,\tau,\nu,\epsilon} \cdot \bS_{i',\tau',\nu',\epsilon})$ is
maintained. As noted above, with the choice of purely imaginary $t_{i,i'}$ we can set
$r_{(i,\tau,\nu),(i',\tau',\nu')}(\epsilon) = \sqrt{|r_{(i,\tau,\nu),(i',\tau',\nu')}(\epsilon)|^2}$
and verify that any further sign in front of the square root does not have any influence on the results.

For energies $\epsilon$ largely exceeding the SOI energy scales, the scattering matrix tends 
to an energy independent quantity, which we cover with the parameter $R_\infty$ as
\begin{equation}
	|r_{(i,\tau,\nu),(i',\tau',\nu')}(\epsilon)|^2
	\sim \frac{R_\infty}{16} 
	(1 + \bS_{i,\tau,\nu} \cdot \bS_{i',\tau',\nu'}).
\label{eq:R_inf}
\end{equation}
Since $R_\infty \propto \Gamma$, we shall use the condition $R_\infty \ll 1$ to 
control the perturbative expansion.

Since the two CNTs cross at an angle $\theta$ the spin directions $\nu$ and $\nu'$, which are
further affected by the magnetic field, are generally not aligned. Therefore, the tunneling interface,
although spin preserving, 
acts as a spin mixer within the \emph{local} spin bases, 
with $\tau-$dependent mixing amplitudes that are tunable through $\epsilon$
and the external magnetic field.
This tunability causes the new features in the noise spectra reported in this paper.

It should finally be stressed that with the Born approximation the scattering matrix,
Eq. \eqref{eq:s}, is no longer unitary, as unitarity imposes identities on inverse
matrices and involves expansions to infinite order. For controlled perturbative expansions,
therefore, the unitarity of the scattering matrix should be used with care.


\section{Normal state conductance and spin-filtering}
\label{sec:cond}

\begin{figure}
   \includegraphics[width=\columnwidth]{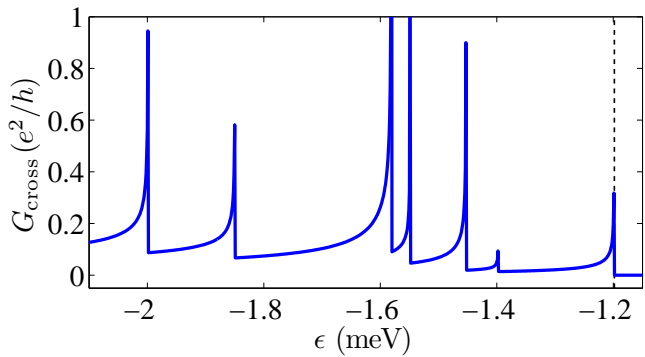}
   \caption{\label{fig:conductance}
   (Color online)
   Cross-conductance $G_\text{cross}$ for a cross-junction of two (30,12) CNTs at 
   an angle $\theta=\pi/3$ in a magnetic field $B=0.6$ T parallel to the outgoing 
   lead $i=4$. Displayed is a zoom on the valence band close to the Dirac points. 
   The peaks indicate the van Hove singularities delimiting the various spin-polarized
   bands in both CNTs. 
   These singularities are rounded off in the numerical implementation.
   The dashed vertical line marks the gap for the tunneling process
   (the larger gap of both CNTs; see the band structure in Fig. \ref{fig:bands}). 
   The tunnel coupling between the CNTs is chosen such that $G_\text{cross} \to (e^2/h) R_\infty$
   at large energies with $R_\infty = 0.05$. 
  }
\end{figure}

The spin-filtering properties of the SOI on the cross-junction first become evident
when considering the normal state cross-conductance. By the Landauer formula, the latter
is given by
\begin{equation}
	G_\text{cross}(\epsilon) = \frac{e^2}{h} \sum_{\tau,\tau',\nu,\nu'} R_{(4,\tau,\nu),(1,\tau',\nu')}(\epsilon),
\end{equation}
with $R_{(4,\tau,\nu),(1,\tau',\nu')} = |r_{(4,\tau,\nu),(1,\tau',\nu')}|^2$. With the definition of $R_\infty$
given in Eq. \eqref{eq:R_inf}, we obtain in the large energy regime $G_\text{cross} \sim (e^2/h) R_\infty$.
An example for this cross-conductance as a function of energy (bias) is shown in Fig. \ref{fig:conductance}.
As mentioned in the previous section, close to the gap the conductance is largely dominated by the
strongly varying densities of states. This leads to the singular peaks in the figure, at which higher
order processes would need to be taken into account. Yet even close to the peaks, the perturbative
expansion remains well controlled. The peak structure indicates that progressively scattering channels
are closed when approaching the gap. Since the SOI causes a spin filtering, this indicates also
that the outgoing current can be spin polarized.

\begin{figure}[!t]
	\includegraphics[width=\columnwidth]{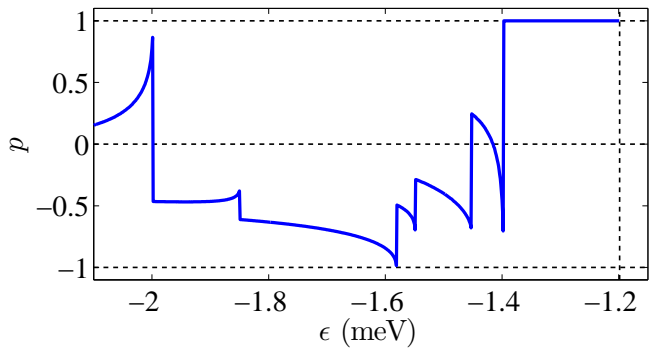}
	\caption{\label{fig:polarization}
   (Color online)
	Spin polarization $p$ for electrons in the outgoing lead $i=4$ for the same conditions 
	as in Fig. \ref{fig:conductance}. Near the gap for the tunneling process (vertical dashed line),
	only one outgoing channel is available and the electrons are perfectly spin polarized.
	Small variations of the gate potential allow effective switching to the opposite spin 
	direction with high (yet not perfect) efficiencies. A reversal of the magnetic field
	$\bB \to -\bB$ reversed all spin directions and can be used for perfect spin filtering
	of the opposite spin direction. The dashed horizontal lines are guides for the eye
	at $p=-1,0,1$.
	}
\end{figure}

To make this evident, we choose the magnetic field parallel to the outgoing lead $i=4$ 
(see Fig. \ref{fig:beamsplitter}),
such that the eigenvalues $\nu = \pm$ coincide for all energies with the spin projections $\up,\dw$
parallel to the CNT axis. The polarization of the outgoing current is then given by
\begin{equation}
	p =
	\frac{
		\sum_{\tau,\tau',\nu'}
		\left[
			R_{(4,\tau,+),(1,\tau',\nu')} - R_{(4,\tau,-),(1,\tau',\nu')}
		\right]
	}{		\sum_{\tau,\tau',\nu'}
		\left[
			R_{(4,\tau,+),(1,\tau',\nu')} + R_{(4,\tau,-),(1,\tau',\nu')}
		\right]
	}.
\end{equation}
A typical result is represented in Fig. \ref{fig:polarization}.
If only one single scattering channel in the outgoing lead $i=4$ is available, full spin polarization
is obtained.
Yet quite remarkably, when crossing with the energy through a band bottom, the strongly enhanced density of
states at the band bottom causes, within a fraction of a meV, a reversal of the
polarization with a final amplitude that can exceed 80\%.
On the other hand, by reversing the magnetic field $\bB \to -\bB$, an
energetically degenerate situation is obtained, yet with switched valleys and spins, such that the full polarization
in the lowest band becomes a full polarization of opposite spin.
This use of the cross-junction as a versatile spin filter at high efficiency complements a previous
suggestion of exploring bound states in SOI-split CNTs for perfect spin 
filtering.\cite{delvalle:2011,jhang:2011}
It also complements the alternative of SOI induced spin-filtering in CNT quantum dot setups.
\cite{braunecker:2013}
%


\section{Injection of entangled electron pairs}
\label{sec:inj}

\begin{figure}
\begin{center}
	\includegraphics[width=\columnwidth]{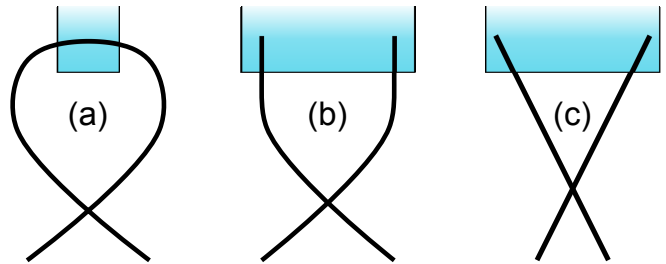}
	\caption{\label{fig:injection_sketches}
   (Color online)
	Possible scenarios for the connection of the CNTs (thick lines) to the entangler (boxes). 
	If the entangler is a 
	superconductor expelling the magnetic field in the parts of the CNTs it covers,
	the spin eigenstates in the CNT parts below the superconductor are parallel to the 
	CNT axes. In situations (a) and (b) the spin eigenvalues in the left and right CNTs
	are then parallel. This allows for an
	\emph{adiabatic} injection, in which any spin correlator of injected electron pairs
	remains identical in the \emph{local} spin bases bound to the CNTs. After adiabatic
	transport to the cross-junction an injected spin singlet, for instance, arrives as
	a spin singlet in the local bases at the cross-junction.
	Situation (c) corresponds to a \emph{nonadiabatic} injection, in which the 
	local spin bases of the left and right CNT are nonparallel below the superconductor
	and the injected pair decomposes into $\nu=\pm$ eigenstates according to the angle
	between the two CNTs below the superconductor. This introduces a deterministic 
	mixture of singlet and triplet states in the local spin bases at the cross-junction.
	}
\end{center}
\end{figure}

We now turn to the injection of spin-entangled electron pairs into the cross-junction,
such as achieved by a Cooper pair splitter.\cite{recher:2001,hermann:2010,schindele:2012} 
The spin-filtering characteristics of the
CNTs require a careful analysis of how the electrons are injected into each CNT,
and how they are transported to the cross-junction. Indeed, if the two CNTs are nonparallel
at the injection region, a first nonparallel spin projection between both CNTs
becomes effective already at injection and affects the entanglement. If the CNTs are
strongly curved in the longitudinal direction, the eigenstates $(\tau,\nu)$ of a straight CNT
can hybridize and the entanglement information of the injected particles can get lost 
(see Fig. \ref{fig:injection_sketches}).

To minimize the latter effect, the energy scale associated with the longitudinal
bending $\epsilon_{lb}$ must be smaller than the sub-meV scales due to the SOI.\cite{flensberg:2010,pei:2012,laird:2012,lai:2012}
Since the bending affects mainly the hopping integral between neighboring carbon ions,
a rough estimate gives $\epsilon_{lb} \sim t R/r$ with $t \sim$ 3 eV the
hopping integral, $R$ the CNT radius, and $r$ the bending radius. A further
reduction of this estimate by averaging over the CNT cross section can be expected.
Since usually $R \sim 1$ nm, therefore, a bending radius larger than
$r \sim 1$ $\mu$m is certainly required.

If such large enough radii can be maintained, the states arriving at the cross-junction
correspond to the states at the injection region.
Figure \ref{fig:injection_sketches}
shows different possible setups of connecting the CNTs to a superconducting entangler. 
In the following, we shall focus mainly on the situation of an \emph{adiabatic} electron
pair injection, in which the spin-correlation state arriving at the cross-junction in 
the \emph{local} spin-eigenbasis $\nu$ bound to each CNT is identical to the 
state in the global spin-eigenbasis in the superconductor. Corresponding possible
setups are shown in Figs. \ref{fig:injection_sketches} (a) and (b). 
We assume furthermore that both CNT branches between the entangler and the cross-junction 
are of comparable length, such that the wave packets of the injected particles strongly
overlap at the cross-junction and the fermion statistics is of importance.
We also focus on small injection rates such that the injected particles have 
a well defined energy spread. 

A \emph{nonadiabatic} situation, in which the local spin bases at injection are
nonparallel and a deterministic singlet-triplet mixture is obtained is displayed
in Fig. \ref{fig:injection_sketches} (c). Its consequences, as well as the consequences
from nonoverlapping wave packets and a larger energy spread, are discussed in 
Sec. \ref{sec:non_adj}.


\section{Current noise}
\label{sec:noise}

Information on the entanglement of the injected electron pairs can be drawn by 
measuring the current noise correlators of the cross-junction. Within the scattering
matrix formalism, the current operator in lead $i$ at time $t$ 
takes the form\cite{buttiker:1992}
\begin{align}
	I^i(t) 
	&= 
	\frac{e}{h}
	\sum_{\substack{j,j',\tau_i,\tau,\tau'\\ \nu_i,\nu,\nu',\epsilon,\epsilon'}}
	A^{i,\tau_i,\nu_i}_{(j,\tau,\nu),(j',\tau',\nu')}(\epsilon,\epsilon') \,
\nonumber \\
	&\times
	a_{j,\tau,\nu}^\dagger(\epsilon) a_{j',\tau',\nu'}(\epsilon')
	\e^{i (\epsilon-\epsilon') t/\hbar},
\end{align}
with 
\begin{gather}
	A^{i,\tau_i,\nu_i}_{(j,\tau,\nu),(j',\tau',\nu')}(\epsilon,\epsilon') 
	=
	\delta_{(i,\tau_i,\nu_i),(j,\tau,\nu)}
	\delta_{(i,\tau_i,\nu_i),(j',\tau',\nu')}
	\delta_{\epsilon,\epsilon'}
\nonumber\\
	-
	s^*_{(i,\tau_i,\nu_i),(j,\tau,\nu)}(\epsilon) 
	s_{(i,\tau_i,\nu_i),(j',\tau',\nu')}(\epsilon').
\label{eq:A_def}
\end{gather}
Since we neglect backscattering into the same lead, the lead indices are restricted
to outgoing leads $i = 3,4$ and incoming leads $j,j' = 1,2$, such that the 
Kronecker symbols in the latter equation drop out.

If $\Ket{\Psi}$ is the state containing the injected particles,
the symmetrized noise correlators for these current operators read
\begin{equation}
	S^{ii'}(t) = \frac{1}{2} \Bra{\Psi}\{\delta I^i(t) , \delta I^{i'}(0)\} \Ket{\Psi},
\end{equation}
with $\delta I^i = I^i - \Bra{\Psi}I^i\Ket{\Psi}$.
The corresponding zero frequency ($\omega=0$) noise is obtained by time 
averaging this quantity as
\begin{equation}
	S^{ii'}
	= \lim_{\mathcal{T}\to \infty}
	\frac{h}{\mathcal{T}} \int_0^{\mathcal{T}} dt \ \mathrm{Re}\, S^{ii'}(t).
\end{equation}
The evaluation of these correlators is carried out in Appendix \ref{app:correlators}. 

Let $\Ket{\Psi}$ describe the injection of two particles, one in lead 1 and one in lead 2,
in either a singlet state $\Ket{\Psi}= \Ket{-}$ or a spin-zero triplet state $\Ket{\Psi} = \Ket{+}$.
If we assume an adiabatic injection as explained in Sec. \ref{sec:inj}, we have
\begin{equation} \label{eq:pm}
	\Ket{\pm} 
	= 
	\sum_{\tau_1,\tau_2,\nu} 
	(\nu)_\pm \
	c_{\tau_1,\tau_2;\nu} \
	a_{2,\tau_2,\nu}^\dagger(\epsilon_0) a_{1,\tau_1,\bar{\nu}}^\dagger(\epsilon_0) 
	\Ket{0},
\end{equation}
with $\Ket{0}$ the equilibrium ground state (we assume temperature $T=0$), and the pair of
particles being injected with certainty into unoccupied states just above the Fermi surface,
such that $\epsilon_0 \gtrsim \epsilon_F$, with $\epsilon_F$ the Fermi energy of the CNT.
We use the convention $\bar{\nu} = -\nu$, and will use it below also for the valley 
indices, $\bar{\tau} = -\tau$.
The symbol $(\nu)_\pm$ defines the signs $(+)_\pm = +$ and $(-)_\pm = \pm$, distinguishing
between triplets and singlets. The wave function amplitudes $c_{\tau_1,\tau_2;\nu}$ are normalized to 
$\sum_{\tau_1,\tau_2,\nu} |c_{\tau_1,\tau_2;\nu}|^2=1$, where it should be noted that only 
those $c_{\tau_1,\tau_2;\nu}$ are nonzero for which any of the pairs of states 
$\{ (1,\tau_1,\nu,\epsilon_0), (2,\tau_2,\bar{\nu},\epsilon_0)\}$ or
$\{ (1,\tau_1,\bar{\nu},\epsilon_0), (2,\tau_2,\nu,\epsilon_0)\}$ has a nonvanishing density of states.
If both pairs exist, the amplitude is independent of $\nu$
to maintain the distinction between singlets and triplets, $c_{\tau_1,\tau_2;\nu} = c_{\tau_1,\tau_2}$. 

In an ideal situation, the only constraint on the energy $\epsilon_0$ is to be larger
than $\epsilon_F$. Realistically, however, a controlled injection of entangled electrons 
requires that $\epsilon_0$ lies close to $\epsilon_F$ to minimize decay processes. 
Yet for optimal operation of the entangler, maintaining still an offset $\epsilon_0-\epsilon_F$
may be favorable.\cite{note}
Exploring the $\epsilon_0$ dependence
of correlators, therefore, corresponds to tuning $\epsilon_F$ through the electron density in the CNTs,
for instance, through a backgate, while maintaining a very small bias $\epsilon_0-\epsilon_F$ for 
the pair injection. 

From Eqs. \eqref{eq:I_j}--\eqref{eq:S_2} we obtain the following result for the cross-correlators 
over the states $\Ket{\pm}$,
assuming henceforth implicit summation over repeated indices (summation within the brackets for the last term),
\begin{align}
	S^{34}_\pm
	&=
	\frac{e^2}{2h}
	c_{\lambda_1,\lambda_2'}^*
	c_{\lambda_1,\lambda_2}
	\bigl[
		A^{3}_{\lambda_2',\gamma}
		A^{4}_{\gamma,\lambda_2}
		+
		A^{4}_{\lambda_2',\gamma}
		A^{3}_{\gamma,\lambda_2}
	\bigr]
\nonumber\\
	&+
	\frac{e^2}{2h}
	c_{\lambda_1',\lambda_2}^*
	c_{\lambda_1,\lambda_2}
	\bigl[
		A^{3}_{\lambda_1',\gamma}
		A^{4}_{\gamma,\lambda_1}
		+
		A^{4}_{\lambda_1',\gamma}
		A^{3}_{\gamma,\lambda_1}
	\bigr]
\nonumber\\
	&+
	\frac{e^2}{h} 
	c_{\lambda_1',\lambda_2'}^* c_{\lambda_1,\lambda_2}
	\Bigl[
		A^{3}_{\lambda_2',\lambda_2} A^{4}_{\lambda_1',\lambda_1}
		-
		A^{3}_{\lambda_2',\lambda_1} A^{4}_{\lambda_1',\lambda_2}
	\Bigr]
\nonumber\\
	&+
	\frac{e^2}{h} 
	c_{\lambda_1',\lambda_2'}^* c_{\lambda_1,\lambda_2}
	\Bigl[
		A^{4}_{\lambda_2',\lambda_2} A^{3}_{\lambda_1',\lambda_1}
		-
		A^{4}_{\lambda_2',\lambda_1} A^{3}_{\lambda_1',\lambda_2}
	\Bigr]
\nonumber\\
	&-
	\frac{e^2}{h}
	\biggl[
		c_{\lambda_1',\lambda_2}^* c_{\lambda_1,\lambda_2}
		A^3_{\lambda_1',\lambda_1}
		+
		c_{\lambda_1,\lambda_2'}^* c_{\lambda_1,\lambda_2}
		A^3_{\lambda_2',\lambda_2}
	\biggr]
\nonumber\\
	&\quad\,\times
	\biggl[
		c_{\lambda_1',\lambda_2}^* c_{\lambda_1,\lambda_2}
		A^4_{\lambda_1',\lambda_1}
		+
		c_{\lambda_1,\lambda_2'}^* c_{\lambda_1,\lambda_2}
		A^4_{\lambda_2',\lambda_2}
	\biggr],
	\label{eq:S_34_full}	
\end{align}
where the $\lambda_i = (i,\tau_i,\nu_i)$, $\lambda_i' = (i,\tau_i',\nu_i')$
are labels bound to lead $i=1,2$,
but $\gamma = (j,\tau,\nu)$ includes the unrestricted summation over $j=1,\dots,4$.
The wave functions are $c_{\lambda_1,\lambda_2} = (\nu_1)_\pm c_{\tau_1,\tau_2;\nu_1} \delta_{\nu_1,\bar{\nu}_2}$,
and we have used the notation 
$A^{3}_{\lambda,\lambda'} = \sum_{\tau,\nu} A^{(3,\tau,\nu)}_{\lambda,\lambda'}(\epsilon_0,\epsilon_0)$.

The first two lines in Eq. \eqref{eq:S_34_full} correspond to single particle like noise, expressed by the quantities
$C_1$ and $S_1$ in Appendix \ref{app:correlators}, which is independent of the type of entanglement between 
the two injected electrons. It turns out that at large energies $\epsilon_0$ these terms vanish.
The following two lines result from the full interference of both electrons
at the junction and are sensitive to the entanglement and the fermion statistics. The last term expresses 
the subtraction of the uncorrelated background current from the product of the currents 
$\Bra{\pm} I^3 \Ket{\pm} \Bra{\pm} I^4 \Ket{\pm}$.

Using $A^3_{\lambda,\lambda'} = - \sum_{\tau,\nu} s_{(3,\tau,\nu),\lambda}^* s_{(3,\tau,\nu),\lambda'}$
and the Born approximation of the scattering matrix, the latter result becomes
\begin{widetext}
\begin{align}
	S^{34}_{\pm}
	&=
	\frac{e^2}{h}
	\mathrm{Re}
	\biggl\{
	c^*_{\tau_1,\tau_2;\nu}
	c_{\tau_1',\tau_2;\nu}
	\Bigl[
		T_{3 \tau_1 \nu}
		r^*_{(4,\tau_4,\nu_4),(1,\tau_1,\nu)}
		r_{(4,\tau_4,\nu_4),(1,\tau_1',\nu)}
		+
		t^*_{3,\tau_1,\nu}
		t^*_{4,\tau_4, \nu_4}
		r_{(3, \tau_1,\nu),   (2,\tau_4,\nu_4)}
		r_{(4, \tau_4,\nu_4), (1,\tau_1', \nu)}
	\Bigr]
\nonumber \\
	+ \
	&
	c^*_{\tau_1,\tau_2;\bar{\nu}}
	c_{\tau_1,\tau_2';\bar{\nu}}
	\Bigl[
		T_{4,\tau_2',\nu}
		r^*_{(3,\tau_3,\nu_3), (2,\tau_2,\nu)}
		r_{(3,\tau_3,\nu_3), (2,\tau_2',\nu)}
		+
		t^*_{3,\tau_3,\nu_3}
		t^*_{4,\tau_2,\nu_2}
		r_{(3,\tau_3,\nu_3),(2,\tau_2',\nu)}
		r_{(4,\tau_2,\nu),(1,\tau_3,\nu_3)}
	\Bigr]
	\biggr\}
\nonumber \\
	+
	\frac{e^2}{h}
	&\biggl\{
		|c_{\tau_1,\tau_2;\nu}|^2
		T_{3,\tau_1,\nu}
		T_{4,\tau_2,\bar{\nu}}
		-
		2 \mathrm{Re}
		\Bigl[
			c_{\tau_1,\tau_2;\nu}^*
			c_{\tau_1',\tau_2';\nu'}
			t^*_{3,\tau_1,\nu}
			t^*_{4,\tau_2,\bar{\nu}}
			r_{(3,\tau_1,\nu),       (2,\tau_2',\bar{\nu}')}
			r_{(4,\tau_2,\bar{\nu}),(1,\tau_1',\nu')}
			\bigl(\delta_{\nu,\nu'}  \pm \delta_{\nu,\bar{\nu}'} \bigr)
		\Bigr]
	\biggr\}
\nonumber \\
	-
	\frac{e^2}{h}
	&\biggl\{
	\Bigl[
		|c_{\tau_1,\tau_2;\nu}|^2 T_{3,\tau_1,\nu}
	\Bigr]
	\Bigl[
		|c_{\tau_1,\tau_2;\nu}|^2 T_{4,\tau_2,\bar{\nu}}
	\Bigr]
	+
	\Bigl[
		|c_{\tau_1,\tau_2;\nu}|^2 T_{3,\tau_1,\nu}
	\Bigr]
	\Bigl[
		c^*_{\tau_1,\tau_2;\nu}
		c_{\tau_1',\tau_2;\nu}
		r^*_{(4,\tau_4,\nu_4), (1,\tau_1,\nu)}
		r_{(4,\tau_4,\nu_4), (1, \tau_1',\nu)}
	\Bigr]
\nonumber \\
	+
	&\ \ \,
	\Bigl[
		|c_{\tau_1,\tau_2;\nu}|^2 T_{4,\tau_2,\bar{\nu}}
	\Bigr]
	\Bigl[
		c^*_{\tau_1,\tau_2;\nu}
		c_{\tau_1,\tau_2';\nu}
		r^*_{(3,\tau_3,\nu_3), (2,\tau_2,\bar{\nu})}
		r_{(3,\tau_3,\nu_3), (2, \tau_2',\bar{\nu})}
	\Bigr]
	\biggr\},
\label{eq:S_34}
\end{align}
\end{widetext}
where the order of terms is the same as in Eq. \eqref{eq:S_34_full},
and where we have used the notations
$t_{3,\tau,\nu} = t_{(3,\tau,\nu),(1,\tau,\nu)}$, 
$t_{4,\tau,\nu} = t_{(4,\tau,\nu),(2,\tau,\nu)}$, 
$T_{3,\tau,\nu} = |t_{3,\tau,\nu}|^2$, 
and
$T_{4,\tau,\nu} = |t_{4,\tau,\nu}|^2$.
All parts of the 
scattering matrix are evaluated at the energy $\epsilon_0$.
For consistency with the Born approximation, we have neglected in the 
latter expression any term on the order of $|r|^4$.

We emphasize that this result for the cross-noise holds for both scattering 
matrices \eqref{eq:s} and \eqref{eq:sp}.
For a scattering matrix of the form of Eq. \eqref{eq:s} we furthermore have
$S^{33}=-S^{34} = - S^{43} = S^{44}$ as a consequence of particle conservation,
independently of the state $\Ket{\Psi}$
(see Appendix \ref{app:identity}). Although the Born approximation violates
the unitarity of the scattering matrix, we have checked by direct comparison 
of the approximate results that $S^{33}_\pm=-S^{34}_\pm$ is indeed maintained.
For a scattering matrix of the form of Eq. \eqref{eq:sp}, however, the latter
equality no longer holds. Indeed, $S^{33}$ acquires then an extra term involving 
the additional scattering between leads $3$ and $4$, such that 
\begin{align}
	S^{33}_\pm 
	&= 
	- S^{34}_\pm
	+ 
	\frac{e^2}{h} 
	c_{\tau_1,\tau_2;\nu}^* 
	c_{\tau_1',\tau_2;\nu}
\nonumber\\
	&\times
	t_{3,\tau_1,\nu}^*
	t_{3,\tau_1',\nu}
	r_{(3,\tau_1',\nu),(4,\tau_4,\nu_4)}^*
	r_{(3,\tau_1,\nu),(4,\tau_4,\nu_4)}.
\end{align}
This extra shift is independent of the entanglement and corresponds to a 
self-energy like renormalization of the single-particle part of the noise correlators
(adding to parts $C_1$ and $S_1$ in Appendix \ref{app:correlators}).
Since such a term tends to obscure the clean signatures of the entanglement, 
we will focus henceforth on the cross-correlators only. 

\begin{figure}
   \includegraphics[width=\columnwidth]{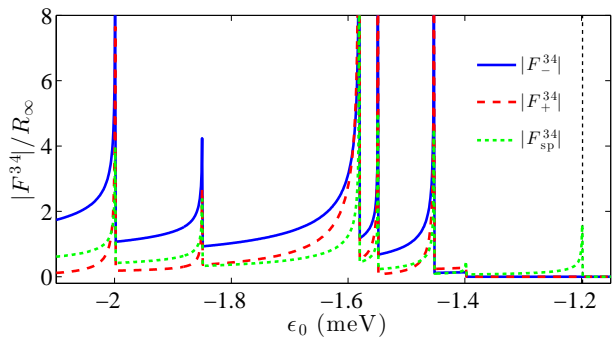}
   \caption{\label{fig:F_34}
   (Color online)
   Fano factors for the triplet and singlet noise $F^{34}_\pm$ and the incoherent
   single-particle noise $F^{34}_\text{sp}$ for the conditions as in Fig. \ref{fig:conductance},
   for the wave function $c_{\tau_1,\tau_2;\nu} \equiv c$, corresponding to an equal valley
   mixing of the injected particles, with $|c|^2 = 1/N$ for $N$ the number of available scattering 
   channels for injection [see the discussion following Eq. \eqref{eq:pm}].
   As explained in the text, the dependence on $\epsilon_0$ reflects usually a dependence on a
   backgate potential.
   The curves are normalized to $R_\infty = 0.05$ [see Eq. \eqref{eq:R_inf}]. The noise correlators show a similar peaked behavior
   as the conductance shown in Fig. \ref{fig:conductance}. All curves are clearly distinct, yet their
   finer structure is better visualized through the modified Fano factors shown in Fig. \ref{fig:F_mod_34}.
   The dashed vertical line marks the gap for the tunnel junction.
  }
 \end{figure}

Since from the unitarity of the scattering matrix it follows that 
$\mean{I^3} = \Bra{\pm} I^3 \Ket{\pm} = e/h$ (not invoking any further Born approximation), 
the Fano factor $F^{34}_\pm = S^{34}_\pm / 2 e \mean{I^3}$ is, up to a constant, the same 
as $S^{34}_\pm$.
In Fig. \ref{fig:F_34} we plot $F^{34}_\pm$ as a function of the energy $\epsilon_0$ of the injected
particles. 
For comparison, we also show the Fano factor resulting
from the incoherent injection of single particles, 
$F_\text{sp}^{34} = S_\text{sp}^{34} / 2 e \mean{I^3+I^4} = S_\text{sp}^{34} / (2 e^2/h)$,
obtained from $S_\text{sp}^{34}(t) = \sum_{i,\nu} \Bra{\Psi_{i,\nu}} \{ \delta I^3(t) , \delta I^4(0) \} \Ket{\Psi_{i,\nu}} / 4$,
for $i=1,2$ with $\Ket{\Psi_{i,\nu}} = \frac{1}{\sqrt{2}} \sum_{\tau} a_{i,\tau,\nu}^\dagger(\epsilon_0)\Ket{0}$.
The corresponding expressions have been derived in Appendix \ref{app:single_particle_inj}.
Note the factor $1/2$ in $F_\text{sp}^{34}$, which causes an identical normalization as for
$F^{34}_\pm$.

\begin{figure}
   \includegraphics[width=\columnwidth]{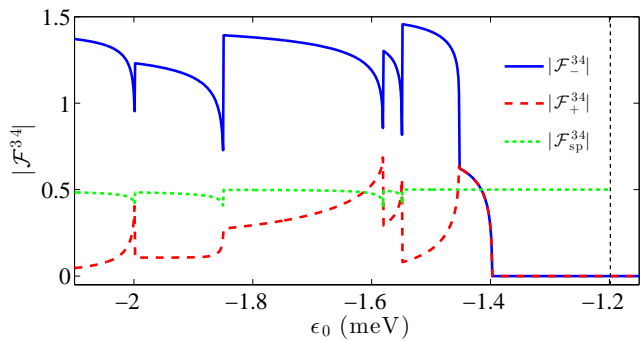}
   \caption{\label{fig:F_mod_34}
   (Color online)
   Modified Fano factors $\mathcal{F}^{34}_\pm$ and $\mathcal{F}^{34}_\text{sp}$
   corresponding to the curves shown in Fig. \ref{fig:F_34}. The dominating peak structure 
   by the van Hove singularities is largely suppressed. Notable distinctions between 
   the triplet, singlet, and uncorrelated single-particle noise correlators are:
   (a) Correlated spin-zero particle pairs have a larger gap in the transport than single particles
   (vertical dashed line) as a result of the spin filtering.
   (b) The asymptotics represent the bunching--antibunching behavior with vanishing
   triplet noise $\mathcal{F}^{34}_+ \to 0$ and a singlet noise $\mathcal{F}^{34}_-$
   exceeding the single particle noise $\mathcal{F}^{34}_\text{sp}$.
   (c) The structure between the gap and the asymptotics is nonuniversal, depending on 
   chirality, $\theta$, and $\bB$, yet is similar for any CNT. The spin filtering 
   effect causes larger variations of the curves for spin-correlated electrons than 
   for the uncorrelated single-particle noise (the value $\mathcal{F}^{34}_{\text{sp}} \approx 0.5$
   is a consequence of the normalization and the averaging of this noise).
  }
 \end{figure}

We observe in Fig. \ref{fig:F_34} that at large energies $|\epsilon_0|$ 
the signature of bunching and antibunching of a structureless conductor, $F^{34}_+ \approx 0$
(yet see Sec. \ref{sec:B} for $B$-field corrections) 
and $F^{34}_- \sim T R$ is recovered.\cite{burkard:2000}
Close to the gap at the Dirac points, however, the Fano factors are 
dominated by the strongly varying densities of states close to the band bottoms, similar to the 
behavior of the cross-conductance. A better resolution of the structure of the correlators
close to the gap is therefore obtained by dividing the noise correlators by the normal state cross-conductance
$G_\text{cross}$, defining a modified Fano factor $\mathcal{F}^{34} = S^{34}/G_\text{cross}$ with
much suppressed singularities at the band bottoms.
Figure \ref{fig:F_mod_34} shows $\mathcal{F}^{34}_\pm$ and $\mathcal{F}^{34}_{\text{sp}}$
as a function of energy.

This figure reveals some remarkable features resulting from the spin-filtering properties. 
We notice first that the gap for spin-correlated pairs is larger than for single-particle transport.
Indeed, if we compare with the conductance in Fig. \ref{fig:conductance} and the polarization in 
Fig. \ref{fig:polarization} we notice that $\mathcal{F}^{34}_\pm = 0$ 
in the full range where $p=1$, and transport
is governed by a single outgoing channel in lead $i=4$, while $\mathcal{F}^{34}_\text{sp}$ remains nonzero. 
Even larger is the gap for the exchange part depending on the $\pm$ sign in the noise expressions, 
and we see that $S^{34}_+ = S^{34}_-$ over a larger energy range. While for large energies $\mathcal{F}^{34}_+ \approx 0$
(up to $B$ dependent corrections; see Sec. \ref{sec:B})
and $\mathcal{F}^{34}_-$ exceeds the single-particle noise, the structure of the curves in the energy range
dominated by the proximity to van Hove singularities strongly depends on the CNT chirality, the
angle $\theta$ between the CNTs, and the magnetic field strength. However, for any choice of the latter
values, a similar shape of the curves is obtained. We notice furthermore that the variations and 
jumps in the correlators
$\mathcal{F}^{34}_\pm$ are more pronounced than those of $\mathcal{F}^{34}_\text{sp}$, which remains 
always close to $\mathcal{F}^{34}_\text{sp} \approx 0.5$. This behavior is closely connected to the
large jumps in polarization $p$ upon varying $\epsilon_0$ (see Fig. \ref{fig:polarization}), which affects
a spin-correlated state much more than an incoherently spin-averaged single-particle state. 
It should also be noted that the value $\mathcal{F}^{34}_\text{sp} \approx 0.5$ results from the 
spin averaging procedure and the chosen normalization of $\mathcal{F}^{34}_\text{sp}$.


\section{Magnetic field dependence}
\label{sec:B}

At energies $|\epsilon|$ far away from all the SOI induced gaps, the spin polarizations
of the various bands become constants, parallel to the effective valley-dependent
fields $\bB_{\tau}^\text{eff} = \tau \bB_{SOI} + \bB$ (we restrict to the lowest
band and drop the band index $n$). Since furthermore the densities of states tend to a constant, 
we see from Eq. \eqref{eq:r2} that the tunneling matrix elements become
\begin{equation} \label{eq:r_asymp}
	|r_{(i,\tau,\nu),(i',\tau',\nu')}|^2 
	\to 
	R \frac{1 + \bS_{i,\tau,\nu} \cdot \bS_{i',\tau',\nu'}}{2},
\end{equation}
with, for convenience, $R = R_\infty/16$ [see Eq. \eqref{eq:R_inf}] 
and spin polarizations independent of $\epsilon$.

To facilitate the discussion, we assume that the magnetic field is applied parallel to 
CNT 2 (leads $i=2,4$, see Fig. \ref{fig:beamsplitter}), 
such that $\bB^\text{eff}_\tau$ is parallel to $\bB$ in this CNT, and
makes an angle $\theta_\tau$ with the other CNT (leads $i=1,3$), 
given by 
\begin{equation}
	\tan(\theta_\tau) = 
	\frac{\tau B_{SOI} \sin(\theta)}{B + \tau B_{SOI} \cos(\theta)},
\end{equation}
with $\theta$ the angle between the CNTs, $B_{SOI} = |\bB_{SOI}|$, and $B = |\bB|$.
Consequently
\begin{align}
	|r_{(3,\tau,\pm),(2,\tau',\pm)}|^2 
	&\to 
	R \frac{1 + \cos(\theta_\tau)}{2}
	= 
	R \cos^2(\theta_\tau/2),
\\
	|r_{(3,\tau,\pm),(2,\tau',\mp)}|^2 
	&\to 
	R \frac{1 + \sin(\theta_\tau)}{2}
	= 
	R \sin^2(\theta_\tau/2).
\end{align}
For CNTs with radii of $\sim 1$ nm, we find that $B_{SOI} \gtrsim 1$ T. 
For external fields up to the tesla range, we then can expand the noise correlators
as a function of $B/B_{SOI} < 1$. With $|t_{(j,\tau,\nu),(1,\tau,\nu)}|^2 \to T$ for $j=3,4$
we obtain from Eq. \eqref{eq:S_34}
\begin{align}
	&S^{34}_{\pm}
	=
	-
	\frac{e^2}{h}
	2 TR \
	\biggl\{
		c_{\tau_1,\tau_2;\nu}^* c_{\tau_1',\tau_2;\nu}
		+
		c_{\tau_1,\tau_2;\nu}^* c_{\tau_1,\tau_2';\nu}
\nonumber\\
		&-
		c_{\tau_1,\tau_2;\nu}^*
		c_{\tau_1',\tau_2';\nu'}
		\Bigl[
			\delta_{\nu,\nu'}  
			\sin^2\frac{\theta}{2}
			\pm 
			\delta_{\nu,\bar{\nu}'} 
			\cos^2\frac{\theta}{2}
		\Bigr]
	\biggr\}
\nonumber \\
	&+
	\frac{e^2}{h}
	TR \frac{\sin^2\theta B^2}{2 B_{SOI}^2}
	\biggl\{
		2 
		c_{\tau_1,\tau_2;\nu}^* c_{\bar{\tau}_1,\tau_2;\nu}
\nonumber \\
	&	+
		c_{\tau_1,\tau_2;\nu}^*
		c_{\tau_1,\tau_2';\nu'}
		3 \cos\theta
		\Bigl[
			\delta_{\nu,\nu'}  
			\mp
			\delta_{\nu,\bar{\nu}'} 
		\Bigr]
\nonumber \\
	&
		+ 
		c_{\tau_1,\tau_2;\nu}^*
		c_{\bar{\tau}_1,\tau_2';\nu'}
		\Bigl[
			\delta_{\nu,\nu'}  
			\bigl( 2 \cos\theta-1 \bigr)
			\mp
			\delta_{\nu,\bar{\nu}'} 
			\bigl( 2 \cos\theta+1 \bigr)
		\Bigr]
	\biggr\}.
\end{align}
If a valley-independent injection is assumed, $c_{\tau_1,\tau_2;\nu} = 1/\sqrt{8}$,
this expansion becomes more transparent. Resolving it explicitly for 
triplet and singlet injection we have, up to quadratic order in $B$,
\begin{align}
	S^{34}_+
	&\sim 
	-
	\frac{e^2}{h} 
	TR \frac{\sin^2(\theta) B^2}{B_{SOI}^2},
	\\
	S^{34}_-
	&\sim 
	-
	\frac{e^2}{h} 
	T R 
	\biggl\{
		16 \cos^2(\theta/2)
		- \frac{\sin^2(\theta) B^2}{B_{SOI}^2}
		\bigl[1 + 10 \cos(\theta)\bigr]
	\biggr\}.
\end{align}
Let us consider first the result at $B=0$. The factor 16 in $S^{34}_-$ is proportional to the number 
of scattering channels and must be compared with the result $S^{34}_- = (e^2/h) 4 T R$ for the single-channel case.\cite{burkard:2000}
Moreover we note the angular dependence on $\cos^2(\theta/2)$ for the singlet
case as a consequence of the spin projections during the tunneling process.

A similar angular dependence is obtained from the SOI in semiconductor beam splitters.\cite{egues:2005} 
Yet in the latter the angle originates from the precession of the spins when traveling through 
a SOI region before reaching the beam splitter and as such is tunable by side gates, while for the CNTs it 
is a consequence of the  crossing angle at the junction and the projective nature of the SOI and is fixed. 

\begin{figure}
   \includegraphics[width=\columnwidth]{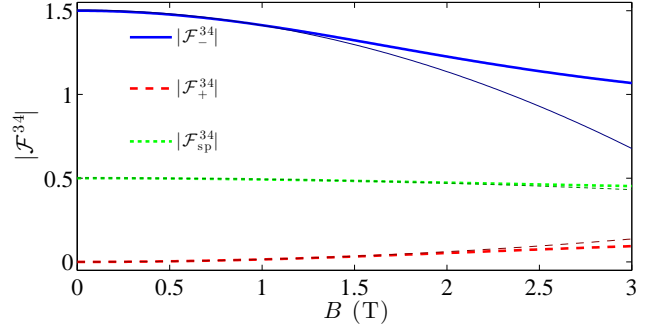}
   \caption{\label{fig:b_field}
   (Color online)
   Magnetic field dependence of the modified Fano factors for the (30,12) CNT cross-junction 
   with $\theta=\pi/3$ as for Fig. \ref{fig:conductance}.
   The energy is fixed at $\epsilon=-30$ meV. The thick curves display the full expressions 
   for $\mathcal{F}^{34}_\pm$ and $\mathcal{F}^{34}_\text{sp}$. The thinner, darker lines
   show the corresponding expansions to order $B^2/B_{SOI}^2$.
     }
 \end{figure}

The main tunability in CNTs arises from the magnetic field dependence.
At nonzero but small $B$, the first $B$-field corrections are quadratic in $B/B_{SOI}$. Remarkably, they 
have opposite signs for singlet and triplet cases, with the triplet increasing from 0 and the singlet 
decreasing from its $B=0$ value. This behavior should be further compared with the $B$-field 
dependence of single particle noise. From the results of Appendix \ref{app:single_particle_inj}
we obtain the expansions, up to order $B^2$, when injecting any spin $\nu$ into lead $i=1,2$,
with equal amplitudes in both valleys $\tau$,
\begin{align}
	S^{34}_{\text{sp},i=1} 
	&\sim
	-
	\frac{e^2}{h} T R \biggl[
		4 
		- 
		\frac{\sin^2(\theta) B^2}{2B_{SOI}^2}
	\biggr],
\label{eq:S_34_sp_1_B_exp}
\\
	S^{34}_{\text{sp},i=2} 
	&\sim
	-
	\frac{e^2}{h} 4 T R.
\end{align}
We note that for particles injected into lead $i=2$ the noise correlators 
are independent of the magnetic field for a scattering matrix of the 
form of Eq. \eqref{eq:r_asymp}, yet a weak dependence can remain through 
the field dependence of the densities of states. 
On the other hand, the $B^2$ 
dependence of the noise correlators for injection into lead $i=1$ 
is similar to the singlet case. Yet for the incoherently averaged single-particle injection 
$S^{34}_{\text{sp}}=( S^{34}_{\text{sp},1} + S^{34}_{\text{sp},2})/2$
the overall amplitude of the $B^2$-dependent terms is much reduced compared 
with $S^{34}_-$. This behavior is visible in Fig. \ref{fig:b_field}, in which we 
display the $B$-field dependence of $\mathcal{F}^{34}_\pm$ and $\mathcal{F}^{34}_{\text{sp}}$.

 %


\section{Influence of deviations from adiabatic injection}
\label{sec:non_adj}

In a realistic implementation, the previous result can be affected by 
several effects that we discuss in the following. Such effects generally 
perturb the clean entanglement signatures, and we provide here estimates of their
influence.

\emph{Nonadiabatic pair injection}. 
In the case of a nonadiabatic spin injection, the electron pair decomposes in 
nonparallel local spin eigenstates at the injection. As a consequence, 
the local triplet and singlet states $\Ket{\pm}$ have further ``local-spin-1'' wave function
components $\Ket{\pm,\pm} \propto a_{1,\tau,\pm}^\dagger a_{2,\tau,\pm}^\dagger \Ket{0}$.
If the injection is spin independent, however, there is no mixing between the local $\Ket{\pm}$ 
states. 
All averages between any of these wave function components are generally nonzero, and  
all the further contributions have a similar form as Eqs. \eqref{eq:S_34_full}
and \eqref{eq:S_34}, with the effect of washing out the derived distinctions 
between the singlet and triplet states. To gauge the amplitude of this perturbation, 
let $\theta_{na}$ be the angle between the spin eigenaxes at injection ($\theta_{na}=0$ 
for adiabatic injection). This angle plays a similar role as $\theta$ for the cross-junction 
and, for instance, for a situation as in Fig. \ref{fig:injection_sketches} (c), $\theta_{na}=\theta$.
Hence the previously derived correlators are weighted by $\cos^2\theta_{na}$, 
and the further terms have amplitudes proportional to $\sin\theta_{na}\cos\theta_{na}$
and $\sin^2\theta_{na}$. For clear entanglement signatures, therefore, $\tan\theta_{na}$ 
should be kept small. 

\emph{Same lead injection}. In any setup, a part of the injected pairs does not split but 
enters the same lead. In a CNT, the Pauli principle is a weaker inhibitor for same-lead injection
than in semiconductor wires, because of the further valley quantum state $\tau$, allowing even 
an equal energy injection. This favors the transfer of spin entanglement onto an orbital entanglement,
and again the resulting correlators have a similar shape as for the split pairs. However, since two 
particles are transmitted within the same CNT, the overall amplitude is proportional to the 
square of the transmission amplitude, $\sim T^2$, and therefore is much larger. Such contributions 
produce a large background on the noise from split electron pairs, which should be detectable
and may be subtracted from the measurement data. 
High splitting efficiencies generally require further interaction effects such 
as the Coulomb blockade.\cite{recher:2001,schindele:2012}

\emph{Different arrival times.} If the two branches of the CNTs between injection points and
the cross-junction are of different length, such that at arrival the wave packets of the two 
injected particles do not overlap, the bunching and antibunching behavior is suppressed.
This corresponds to the case in which the terms $C_2$ and $S_2$ in Eqs. \eqref{eq:C_2} and \eqref{eq:S_2}
vanish. The averaged noise then maintains some information on the spin correlations, but the
information on the entanglement arising from the exchange part of the correlators proportional to the 
$\pm$ signs is lost. 

\emph{Energy spread of the injected wave packets.}
It was assumed in the calculation that the energy levels of the injected particles are 
fixed to $\epsilon_0$. This requires the limit of low temperatures $T$ and small tunnel rates 
for the injection $\Gamma_{\text{inj}}$, such that $k_B T, \Gamma_{\text{inj}} < \Delta \epsilon$,
with $\Delta \epsilon$ the level spacing of the CNTs. While maintaining $k_B T < \Delta \epsilon$
is desirable to avoid covering the signal behind the thermal noise $\sim T$, the smallness of 
$\Gamma_{\text{inj}}$ is less critical if the scattering at the cross-junction is 
elastic. In this case, all displayed curves are just smeared out over energy windows of the width 
$\Gamma_{\text{inj}}$, centered about the values $\epsilon_0$. This is different from  
a mesoscopic beam splitter that allows for inelastic processes, for which $\Gamma_{\text{inj}} > \Delta\epsilon$
has a larger impact.\cite{egues:2005}

\emph{Valley-selective tunneling.} In the shown figures for the noise correlators
we have considered wave function amplitudes $c_{\tau_1,\tau_2;\nu} \equiv c$ that contained
an equal distribution of the injected electron over the two valleys. Such an equal distribution
is obtained when the tunneling into the CNTs is mostly local. Different contacts, however, are 
perfectly possible: for instance, some specific $\tau$ bound injection into a CNT, or the situation 
in which the opposite momenta of Cooper pairs in the superconductor are maintained in the form
of tunneling into opposite valleys $(i=1,\tau)$ and $(i=2,\bar{\tau})$ only. 
Such situations impose further constraints on the correlators, and overall just reduce their amplitude,
yet do not affect them qualitatively. More subtle is the case, in which the wave functions in different
valleys pick up different (deterministic or random) phases during transport to the cross-junction. This leads 
to an orbital interference
effect that competes with the singlet--triplet signatures, and as a general rule produces correlators 
that lie somewhere between the singlet and triplet results.


\section{Conclusions}
\label{sec:conclusions}

In this paper we have shown that CNT cross-junctions have rich and tunable spin dependent properties
due to the SOI. 
First, this turns such cross-junctions into versatile spin filters, allowing
generically to obtain perfect spin polarizations at low energies. By reversing the magnetic field
$\bB \to -\bB$, the polarization is reversed as well. Opposite polarizations
that can exceed 80\% are also achievable by small changes of the backgate potential.

Second, the SOI adds further possibilities to obtain information on the entanglement of injected
pairs of spins. In particular, it strongly affects the bunching and antibunching behavior of
injected singlet and triplet states. At energies in which the number of scattering channels is 
reduced, the spin projective properties allow us to distinguish between spin correlated and
spin uncorrelated pairs, notably with correlated pairs of opposite spin having a larger gap in transport.
At larger energies, at which the densities of state become constant, the spin exchange parts of the 
noise correlators acquire a dependence on $\theta$ similar to the dependence 
on the SOI-angle in the semiconductor setup described in Ref. \onlinecite{egues:2005}. However,
for the CNTs the angle $\theta$, which is the crossing angle of the junction, is not tunable, in contrast to 
the semiconductor case. Tunable instead is the magnetic field $\bB$, and we have shown that singlet and 
triplet correlations lead to a qualitatively different $\bB$ dependence, clearly distinct from the 
dependence of noise of uncorrelated particles.

The described effects are pronounced in all types of CNTs, with the exception of
armchair CNTs which do not have the gap induced by SOI and curvature.

To conclude we note that the use of ferromagnetic contacts on the outgoing leads could
provide further indicators for entanglement. Indeed, the most sensitive part of the noise
correlators to magnetic filtering are the exchange terms distinguishing between singlets 
and triplets. By rotating, for instance, the magnetization direction of the contacts through
various angles, we expect therefore a modulation of the noise correlators with opposite sign 
for the singlet and triplet cases, on top of the behavior investigated in this paper. 
A systematic study of this effect is left for future work.


\acknowledgments
We thank A. Baumgartner, Y. Blanter, C. Bruder, R. Fazio, V. Giovannetti, J. Schindele, 
C. Sch\"{o}nenberger, and F. Taddei for helpful discussions. 
This work has been supported by the EU-FP7 project SE2ND [271554], 
by the Italian MIUR-PRIN,
and by the Spanish MINECO through Grant No. FIS2011-26516, and by the DFG
Grant No. RE 2978/1-1.


\appendix


\section{Evaluation of the correlators}
\label{app:correlators}

\subsection{General form of correlators}

The noise correlators are obtained by evaluating expectation values of the
general form
\begin{equation}
	C(t) =
	\e^{i (\epsilon_{\gamma}-\epsilon_{\gamma'})t/\hbar}
	\Bra{\lambda_1,\lambda_2}
	a_{\gamma}^\dagger a_{\gamma'}
	a_{\delta}^\dagger a_{\delta'}
	\Ket{\lambda_1',\lambda_2'},
\end{equation}
with the labels $\lambda,\gamma,\dots$ covering all internal quantum states for
the scattering states, including leads, valleys, and spin eigenstates, and where
we have used $a_\gamma(t) = e^{-i \epsilon_\gamma t/\hbar} a_\gamma$.
The time averaging done by a measurement is expressed by the integral
\begin{equation}
	C 
	= \frac{h}{\mathcal{T}} \int_{0}^{\mathcal{T}} dt \ \mathrm{Re}\, C(t)
	= \frac{h}{2\mathcal{T}} \int_{-\mathcal{T}}^{\mathcal{T}} dt \ C(t),
\end{equation}
in which we let $\mathcal{T} \to \infty$. Through
the normalization by the Planck constant $h$ the resulting noise has the
dimension of $e^2/h$.
Since the only time-dependent quantity in $C(t)$ is the phase factor, the time averaging
gives a factor $\delta_{\epsilon_{\gamma},\epsilon_{\gamma'}}$, such that
\begin{equation}
	C =
	h \delta_{\epsilon_{\gamma},\epsilon_{\gamma'}}
	\Bra{\lambda_1,\lambda_2}
	a_{\gamma}^\dagger a_{\gamma'}
	a_{\delta}^\dagger a_{\delta'}
	\Ket{\lambda_1',\lambda_2'},
\end{equation}
By Wick's theorem we can write $C = h \delta_{\epsilon_{\gamma},\epsilon_{\gamma'}} \sum_{n=0}^2 C_n$, where $n$
denotes the number of contractions between the operators of the injected particles $\lambda_{1,2}$
and the operators related to the current operator, $\gamma,\gamma',\delta,\delta'$.
Since $\Ket{\lambda_1,\lambda_2} = a_{\lambda_1}^\dagger a_{\lambda_2}^\dagger \Ket{0}$,
we have
\begin{equation}
	C_0 =
	(
		\delta_{\lambda_1,\lambda_1'} \delta_{\lambda_2,\lambda_2'}
		+
		\delta_{\lambda_1,\lambda_2'} \delta_{\lambda_2,\lambda_1'}
	)
	\Bra{0} a_{\gamma}^\dagger a_{\gamma'} a_{\delta}^\dagger a_{\delta'} \Ket{0},
\end{equation}
in which the remaining average corresponds to the usual noise at thermal equilibrium.
In the low-temperature limit $k_B T < \Delta \epsilon$, with $\Delta\epsilon$ the
level spacing, these contributions vanish. Indeed,
\begin{align}
	&\Bra{0} a_{\gamma}^\dagger a_{\gamma'} a_{\delta}^\dagger a_{\delta'} \Ket{0}
\nonumber\\
	&=
	\delta_{\gamma,\gamma'} \delta_{\delta,\delta'} f(\epsilon_\gamma) f(\epsilon_\delta)
	+
	\delta_{\gamma,\delta'} \delta_{\delta,\gamma'} [1-f(\epsilon_\gamma)] f(\epsilon_\delta),
\end{align}
where $f(\epsilon)$ is the Fermi function.
The first term corresponds to product of current averages and eventually will be subtracted in the
noise correlators. For elastic processes we have $\epsilon_\gamma=\epsilon_\delta$ and
the second term vanishes because
$f(\epsilon_\gamma)[1-f(\epsilon_\gamma)] = 0$ at low temperatures for discrete levels.
For $k_B T > \Delta\epsilon$, it would produce the usual $\sim T$ dependence
of thermal noise.

Since the injected particles are placed above the Fermi surface with certainty, their contractions
are not weighted by the Fermi distribution and are equal to 1.
Therefore, the contribution including one contraction with the injected particles is
\begin{align}
	C_1
	&=
	\delta_{\lambda_1,\lambda_1'}
	\Bra{\lambda_2} a_{\gamma}^\dagger a_{\gamma'} a_{\delta}^\dagger a_{\delta'} \Ket{\lambda_2'}_{\text{conn}}
\nonumber\\
	&+
	\delta_{\lambda_1,\lambda_2'}
	\Bra{\lambda_2} a_{\gamma}^\dagger a_{\gamma'} a_{\delta}^\dagger a_{\delta'} \Ket{\lambda_1'}_{\text{conn}}
\nonumber\\
	&+
	\delta_{\lambda_2,\lambda_1'}
	\Bra{\lambda_1} a_{\gamma}^\dagger a_{\gamma'} a_{\delta}^\dagger a_{\delta'} \Ket{\lambda_2'}_{\text{conn}}
\nonumber\\
	&+
	\delta_{\lambda_2,\lambda_2'}
	\Bra{\lambda_1} a_{\gamma}^\dagger a_{\gamma'} a_{\delta}^\dagger a_{\delta'} \Ket{\lambda_1'}_{\text{conn}},
\end{align}
in which the fully connected correlators of a single injected particle are given by
\begin{align}
	&\Bra{\lambda} a_{\gamma}^\dagger a_{\gamma'} a_{\delta}^\dagger a_{\delta'} \Ket{\lambda'}_{\text{conn}}
\nonumber\\
	&=
	\contraction{}{a}{{}_{\lambda}}{a}
	\contraction{a_{\lambda} a_{\gamma}^\dagger}{a}{{}_{\gamma'}}{a}
	\contraction{a_{\lambda} a_{\gamma}^\dagger a_{\gamma'} a_{\delta}^\dagger}{a}{{}_{\delta'}}{a}
	a_{\lambda} a_{\gamma}^\dagger a_{\gamma'} a_{\delta}^\dagger a_{\delta'} a_{\lambda'}^\dagger
	+
	\contraction{}{a}{{}_{\lambda}}{a}
	\contraction[1.5ex]{a_{\lambda} a_{\gamma}^\dagger}{a}{{}_{\gamma'} a_{\delta}^\dagger a_{\delta'}}{a}
	\contraction{a_{\lambda} a_{\gamma}^\dagger a_{\gamma'}}{a}{{}_{\delta}^\dagger}{a}
	a_{\lambda} a_{\gamma}^\dagger a_{\gamma'} a_{\delta}^\dagger a_{\delta'} a_{\lambda'}^\dagger
\nonumber\\
	&+
	\contraction[1.5ex]{}{a}{{}_{\lambda} a_{\gamma}^\dagger a_{\gamma'}}{a}
	\contraction{a_{\lambda}}{a}{{}_{\gamma}^\dagger}{a}
	\contraction{a_{\lambda} a_{\gamma}^\dagger a_{\gamma'} a_{\delta}^\dagger}{a}{{}_{\delta'}}{a}
	a_{\lambda} a_{\gamma}^\dagger a_{\gamma'} a_{\delta}^\dagger a_{\delta'} a_{\lambda'}^\dagger
	+
	\contraction[1.5ex]{}{a}{{}_{\lambda} a_{\gamma}^\dagger a_{\gamma'}}{a}
	\contraction[2ex]{a_{\lambda}}{a}{{}_{\gamma}^\dagger a_{\gamma'} a_{\delta}^\dagger}{a}
	\contraction{a_{\lambda} a_{\gamma}^\dagger}{a}{{}_{\gamma'} a_{\delta}^\dagger a_{\delta'}}{a}
	a_{\lambda} a_{\gamma}^\dagger a_{\gamma'} a_{\delta}^\dagger a_{\delta'} a_{\lambda'}^\dagger
\nonumber\\
	&=
	\delta_{\lambda,\gamma} \delta_{\gamma',\delta} \delta_{\delta',\lambda'} [1-f(\epsilon_\gamma)]
	+
	\delta_{\lambda,\gamma} \delta_{\gamma',\lambda'} \delta_{\delta',\delta} f(\epsilon_\delta)
\nonumber\\
	&+
	\delta_{\lambda,\delta} \delta_{\gamma',\gamma} \delta_{\delta',\lambda'} f(\epsilon_\gamma)
	-
	\delta_{\lambda,\delta} \delta_{\gamma',\lambda'} \delta_{\delta',\gamma} f(\epsilon_\delta).
\end{align}
Since low temperatures $k_B T < \Delta \epsilon$ and elastic scattering are considered, all the Fermi functions
in the latter equation vanish, as their energy is pinned to the energy of the injected particles above the Fermi surface.
The remaining term gives the result
\begin{align}
	\Bra{\lambda} a_{\gamma}^\dagger a_{\gamma'} a_{\delta}^\dagger a_{\delta'} \Ket{\lambda'}_{\text{conn}}
	=
	\delta_{\lambda,\gamma} \delta_{\gamma',\delta} \delta_{\delta',\lambda'}.
\end{align}
Finally, the fully contracted part reads
\begin{align}
	C_2 &=
	\contraction[1.5ex]{}{a}{{}_{\lambda_2} a_{\lambda_1} a_{\gamma}^\dagger a_{\gamma'}}{a}
	\contraction{a_{\lambda_2}}{a}{{}_{\lambda_1}}{a}
	\contraction{a_{\lambda_2} a_{\lambda_1} a_{\gamma}^\dagger}{a}{{}_{\gamma'} a_{\delta}^\dagger a_{\delta'}}{a}
	\contraction[1.5ex]{a_{\lambda_2} a_{\lambda_1} a_{\gamma}^\dagger a_{\gamma'} a_{\delta}^\dagger}{a}{{}_{\delta'} a_{\lambda_1'}^\dagger}{a}
	a_{\lambda_2} a_{\lambda_1} a_{\gamma}^\dagger a_{\gamma'} a_{\delta}^\dagger a_{\delta'} a_{\lambda_1'}^\dagger a_{\lambda_2'}^\dagger
	+
	\contraction[1.5ex]{}{a}{{}_{\lambda_2} a_{\lambda_1}}{a}
	\contraction{a_{\lambda_2}}{a}{{}_{\lambda_1} a_{\gamma}^\dagger a_{\gamma'}}{a}
	\contraction[1.5ex]{a_{\lambda_2} a_{\lambda_1} a_{\gamma}^\dagger}{a}{{}_{\gamma'} a_{\delta}^\dagger a_{\delta'}}{a}
	\contraction{a_{\lambda_2} a_{\lambda_1} a_{\gamma}^\dagger a_{\gamma'} a_{\delta}^\dagger}{a}{{}_{\delta'} a_{\lambda_1'}^\dagger}{a}
	a_{\lambda_2} a_{\lambda_1} a_{\gamma}^\dagger a_{\gamma'} a_{\delta}^\dagger a_{\delta'} a_{\lambda_1'}^\dagger a_{\lambda_2'}^\dagger
\nonumber\\
	&+
	\contraction[1.5ex]{}{a}{{}_{\lambda_2} a_{\lambda_1}}{a}
	\contraction{a_{\lambda_2}}{a}{{}_{\lambda_1} a_{\gamma}^\dagger a_{\gamma'}}{a}
	\contraction[1.5ex]{a_{\lambda_2} a_{\lambda_1} a_{\gamma}^\dagger}{a}{{}_{\gamma'} a_{\delta}^\dagger a_{\delta'} a_{\lambda_1'}^\dagger}{a}
	\contraction{a_{\lambda_2} a_{\lambda_1} a_{\gamma}^\dagger a_{\gamma'} a_{\delta}^\dagger}{a}{{}_{\delta'}}{a}
	a_{\lambda_2} a_{\lambda_1} a_{\gamma}^\dagger a_{\gamma'} a_{\delta}^\dagger a_{\delta'} a_{\lambda_1'}^\dagger a_{\lambda_2'}^\dagger
	+
	\contraction[1.5ex]{}{a}{{}_{\lambda_2} a_{\lambda_1} a_{\gamma}^\dagger a_{\gamma'}}{a}
	\contraction{a_{\lambda_2}}{a}{{}_{\lambda_1}}{a}
	\contraction{a_{\lambda_2} a_{\lambda_1} a_{\gamma}^\dagger}{a}{{}_{\gamma'} a_{\delta}^\dagger a_{\delta'} a_{\lambda_1'}^\dagger}{a}
	\contraction[1.5ex]{a_{\lambda_2} a_{\lambda_1} a_{\gamma}^\dagger a_{\gamma'} a_{\delta}^\dagger}{a}{{}_{\delta'}}{a}
	a_{\lambda_2} a_{\lambda_1} a_{\gamma}^\dagger a_{\gamma'} a_{\delta}^\dagger a_{\delta'} a_{\lambda_1'}^\dagger a_{\lambda_2'}^\dagger
\nonumber\\
	&=
	\delta_{\lambda_1,\gamma}
	\delta_{\lambda_2,\delta}
	\delta_{\lambda_1', \gamma'}
	\delta_{\lambda_2', \delta'}
	-
	\delta_{\lambda_1,\delta}
	\delta_{\lambda_2,\gamma}
	\delta_{\lambda_1', \gamma'}
	\delta_{\lambda_2', \delta'}
\nonumber\\
	&+
	\delta_{\lambda_1,\delta}
	\delta_{\lambda_2,\gamma}
	\delta_{\lambda_1', \delta'}
	\delta_{\lambda_2', \gamma'}
	-
	\delta_{\lambda_1,\gamma}
	\delta_{\lambda_2,\delta}
	\delta_{\lambda_1', \delta'}
	\delta_{\lambda_2', \gamma'}.
\label{eq:C_2}
\end{align}
By a similar investigation we obtain the expectation values for the current operator
\begin{align}
	I
	&= \Bra{\lambda_1,\lambda_2} a_{\gamma}^\dagger a_{\gamma'} \Ket{\lambda_1',\lambda_2'}
\nonumber\\
	&= I_{eq}
	+ \delta_{\lambda_1,\lambda_1'} \delta_{\lambda_2,\gamma} \delta_{\gamma',\lambda_2'}
	+ \delta_{\lambda_1,\lambda_2'} \delta_{\lambda_2,\gamma} \delta_{\gamma',\lambda_1'}
\nonumber\\
	&
	+ \delta_{\lambda_2,\lambda_1'} \delta_{\lambda_1,\gamma} \delta_{\gamma',\lambda_2'}
	+ \delta_{\lambda_2,\lambda_2'} \delta_{\lambda_1,\gamma} \delta_{\gamma',\lambda_1'},
\end{align}
where $I_{eq}$ contributes to the background current that 
is independent of the injected particles
and vanishes in the zero-bias, low-temperature limit considered here.

\subsection{Pair injection in two leads}

The situation of injecting each particle in a different CNT is expressed by a wave function
$\Ket{\Psi} = \sum_{\lambda_1,\lambda_2} c_{\lambda_1,\lambda_2} \Ket{\lambda_1,\lambda_2}$
with $\lambda_i$ restricted to lead $i=1,2$ and $c_{\lambda_1,\lambda_2}$ the wave function
amplitudes. From the latter results, the current in lead $j$ is
\begin{align}
	\mean{I^j}
	=
	\Bra{\Psi} I^j \Ket{\Psi} =
	&\frac{e}{h}\sum_{\lambda_1,\lambda_1',\lambda_2}
	c_{\lambda_1,\lambda_2}^* c_{\lambda_1',\lambda_2}
	A^j_{\lambda_1,\lambda_1'}
\nonumber \\
	+
	&\frac{e}{h}\sum_{\lambda_1,\lambda_2,\lambda_2'}
	c_{\lambda_1,\lambda_2}^* c_{\lambda_1,\lambda_2'}
	A^j_{\lambda_2,\lambda_2'},
\label{eq:I_j}
\end{align}
with $A^j_{\lambda,\lambda'} = \sum_{\tau,\nu} A^{j,\tau,\nu}_{\lambda,\lambda'}$.
The noise correlators, on the other hand, are
\begin{equation}
	S^{jj'}
	=
	\frac{1}{2}\Bra{\Psi}\{\delta I^j, \delta I^{j'}\}\Ket{\Psi}
	= \frac{e^2}{h} \bigl( S_1 + S_2 \bigr) - h \mean{I^j} \mean{I^{j'}},
\label{eq:S_jj'}
\end{equation}
with
\begin{align}
	S_1
	&=
	\frac{1}{2}
	\sum_{\lambda_1,\lambda_2,\lambda_2',\gamma}
	c_{\lambda_1,\lambda_2}^*
	c_{\lambda_1,\lambda_2'}
	\bigl[
		A^{j}_{\lambda_2,\gamma}
		A^{j'}_{\gamma,\lambda_2'}
		+
		A^{j'}_{\lambda_2,\gamma}
		A^{j}_{\gamma,\lambda_2'}
	\bigr]
\nonumber\\
	&+
	\frac{1}{2}
	\sum_{\lambda_1,\lambda_1',\lambda_2,\gamma}
	c_{\lambda_1,\lambda_2}^*
	c_{\lambda_1',\lambda_2}
	\bigl[
		A^{j}_{\lambda_1,\gamma}
		A^{j'}_{\gamma,\lambda_1'}
		+
		A^{j'}_{\lambda_1,\gamma}
		A^{j}_{\gamma,\lambda_1'}
	\bigr],
\label{eq:S_1}
\end{align}
and
\begin{align}
	S_2
	=
	\sum_{\lambda_1,\lambda_1',\lambda_2,\lambda_2'}
	&c_{\lambda_1,\lambda_2}^* c_{\lambda_1',\lambda_2'}
	\Bigl[
		A^{j}_{\lambda_2,\lambda_2'} A^{j'}_{\lambda_1,\lambda_1'}
		-
		A^{j}_{\lambda_2,\lambda_1'} A^{j'}_{\lambda_1,\lambda_2'}
	\Bigr]
\nonumber \\
	&+
	(j \leftrightarrow j').
\label{eq:S_2}
\end{align}

\subsection{Single particle injection}
\label{app:single_particle_inj}

The noise of a single particle that is injected into lead $i=1,2$ is 
determined by averaging over the wave function
$\Ket{\Psi_{i}} = \sum_{\tau,\nu} c_{i,\tau,\nu} a_{i,\tau,\nu}^\dagger(\epsilon_0) \Ket{0}$,
for $\sum_{\tau,\nu} |c_{i,\tau,\nu}|^2 = 1$. 
The noise correlators for this state are straightforwardly obtained from the 
previous results by setting $C_2 = 0$ and by retaining only the term proportional
to $\Bra{\lambda_i'} \dots \Ket{\lambda_i}$ in $C_1$. 
It follows that
\begin{align}
	&S^{jj'}_{\text{sp},i}
	= 
	\frac{e^2}{2h} \sum_{\lambda_i,\lambda_i',\gamma}
	c_{\lambda_i}^* c_{\lambda_i'} 
	\bigl[
		A^{j}_{\lambda_i,\gamma} A^{j'}_{\gamma,\lambda_i'}
		+
		A^{j'}_{\lambda_i,\gamma} A^{j}_{\gamma,\lambda_i'}
	\bigr]
\nonumber \\
	&
	- \frac{e^2}{h}
	\biggl(
		\sum_{\lambda_i,\lambda_i'} 
		c_{\lambda_i}^* c_{\lambda_i'} 
		A^{j}_{\lambda_i,\lambda_i'} 
	\biggr)
	\biggl(
		\sum_{\lambda_i,\lambda_i'} 
		c_{\lambda_i}^* c_{\lambda_i'} 
		A^{j'}_{\lambda_i,\lambda_i'} 
	\biggr),
\end{align}
with indices $\lambda_i,\lambda_i'$ bound to lead $i$.

Within the Born approximation we obtain for the cross-correlators,
with implicit summation over all indices (summation within each bracket 
in both second lines),
\begin{widetext}
\begin{align}
	S^{34}_{\text{sp},1}
	&= \frac{e^2}{h} 
	\mathrm{Re}
	\biggl\{
		c_{1,\tau_1, \nu}^* 
		c_{1,\tau_1',\nu}
		\Bigl[
			T_{3,\tau_1,\nu}
			r^*_{(4,\tau_4,\nu_4), (1,\tau_1,\nu)}
			r_{(4,\tau_4,\nu_4), (1,\tau_1', \nu)}
			+
			t^*_{(3,\tau_1,\nu)}
			t^*_{(4,\tau_2,\nu_2)}
			r_{(3,\tau_1,\nu),(2,\tau_2,\nu_2)}
			r_{(4,\tau_2,\nu_2),(1,\tau_1',\nu)}
		\Bigr]
	\biggr\}
\nonumber \\
	&-
	\frac{e^2}{h}
	\Bigl[
		|c_{1,\tau_1,\nu}|^2
		T_{3,\tau_1,\nu}
	\Bigr]
	\Bigl[
		c_{1,\tau_1,\nu}^*
		c_{1,\tau_1',\nu}
		r^*_{(4,\tau_4,\nu_4), (1, \tau_1, \nu)}
		r_{(4,\tau_4,\nu_4), (1,\tau_1' \nu)}
	\biggr],
\\
	S^{34}_{\text{sp},2}
	&=
	\frac{e^2}{h}
	\mathrm{Re}
	\biggl\{
		c_{2,\tau_2,\nu}^*
		c_{2,\tau_2',\nu}
		\Bigl[
			T_{4,\tau_2',\nu}
			r^*_{(3,\tau_3,\nu_3), (2,\tau_2,\nu)}
			r_{(3,\tau_3,\nu_3),(2,\tau_2',\nu)}
			+
			t^*_{3,\tau_3,\nu_3}
			t^*_{4,\tau_1,\nu}
			r_{(3,\tau_3,\nu_3), (2,\tau_2',\nu)}
			r_{(4,\tau_2,\nu),(1,\tau_3,\nu_3)}
		\Bigr]
	\biggl\}
\nonumber \\
	&-
	\frac{e^2}{h}
	\Bigl[
		|c_{2,\tau_2,\nu}|^2
		T_{4,\tau_2, \nu}
	\Bigr]
	\Bigl[
		c_{2,\tau_2,\nu}^*
		c_{2,\tau_2',\nu}
		r^*_{(3,\tau_3,\nu_3),(2,\tau_2,\nu)}
		r_{(3,\tau_3,\nu_3),(2,\tau_2',\nu)}
	\Bigr],
\end{align}
\end{widetext}
$t_{3,\tau,\nu} = t_{(3,\tau,\nu),(1,\tau,\nu)}$, 
$t_{4,\tau,\nu} = t_{(4,\tau,\nu),(2,\tau,\nu)}$, 
$T_{3,\tau,\nu} = |t_{3,\tau,\nu}|^2$, 
and
$T_{4,\tau,\nu} = |t_{4,\tau,\nu}|^2$.

Specializing, for instance, to the wave functions $\Ket{\Psi_{i,\nu_0}}$ defined by the 
amplitudes $c_{i,\tau,\nu} = \delta_{\nu,\nu_0} / \sqrt{2}$ allows us to 
capture the injection of a spin $\nu_0$ into lead $i$ with equal weights in both valleys.


\section{Demonstration that $S^{33}=-S^{34}$ in the absence of backscattering}
\label{app:identity}

In the absence of backscattering into the incoming leads $i=1,2$, described by a scattering 
matrix as in Eq. \eqref{eq:s},
any injected particles are fully transmitted into the outgoing leads $i=3,4$.
Hence, by particle conservation (the unitarity of the scattering matrix),
$I^3+I^4 = - \hat{N}_\text{in}$, where 
$\hat{N}_\text{in} = \sum_{i=1,2,\tau,\nu,\epsilon} a_{i,\tau,\nu,\epsilon}^\dagger a_{i,\tau,\nu,\epsilon}$ counts the number of incoming particles. Therefore, for any 
state $\Ket{\Psi}$ with a number $N$ of incoming particles, $\hat{N}_\text{in} \Ket{\Psi} = N \Ket{\Psi}$
and
$\Bra{\Psi} \{ I^3 , (I^3+I^4) \} \Ket{\Psi} = -2 N \Bra{\Psi} I^3 \Ket{\Psi}
= 2 \Bra{\Psi} I^3 \Ket{\Psi} \Bra{\Psi}(I^3+I^4)\Ket{\Psi}$.
Consequently 
$S^{33}+S^{34} = \frac{1}{2} \Bra{\Psi} \{ I^3 , (I^3+I^4) \} \Ket{\Psi} - \Bra{\Psi} I^3 \Ket{\Psi} \Bra{\Psi}(I^3+I^4)\Ket{\Psi} = 0$.
By precisely the same argument $S^{44}+S^{34} = 0$, and therefore $S^{33} = S^{44}$.

These identities are independent of whether $\Ket{\Psi}$ is entangled or not, but the provided
proof depends on the absence of backscattering in the scattering matrix, Eq. \eqref{eq:s}.

\vfill



\begin{thebibliography}{88}

\bibitem{tans:1997}
S. J. Tans, M. H. Devoret, H. Dai, A. Thess, R. E. Smalley, L. J. Geerligs, and C. Dekker, 
Nature (London) \textbf{386}, 474 (1997).

\bibitem{cobden:2002}
D. H. Cobden and J. Nyg{\aa}rd, 
Phys. Rev. Lett. \textbf{89}, 046803 (2002).

\bibitem{liang:2002}
W. Liang, M. Bockrath, and H. Park, 
Phys. Rev. Lett. \textbf{88}, 126801 (2002).

\bibitem{minot:2004}
E. Minot, Y. Yaish, V. Sazonova, and P. McEuen, 
Nature (London) \textbf{428}, 536 (2004).

\bibitem{jarillo-herrero:2004}
P. Jarillo-Herrero, S. Sapmaz, C. Dekker, L. P. Kouwenhoven, and H. S. J. van der Zant, 
Nature (London) \textbf{429}, 389 (2004).

\bibitem{jarillo-herrero:2005}
P. Jarillo-Herrero, J. Kong, H. S. J. van der Zant, C. Dekker, L. P. Kouwenhoven, and S. De Franceschi, 
Phys. Rev. Lett. \textbf{94}, 156802 (2005).

\bibitem{graeber:2006}
M. R. Gr\"{a}ber, W. A. Coish, C. Hoffmann, M. Weiss, J. Furer,
S. Oberholzer, D. Loss, and C. Sch\"{o}nenberger, 
Phys. Rev. B \textbf{74}, 075427 (2006).

\bibitem{vijayaraghavan:2012}
A. Vijayaraghavan,
J. Mater. Chem. \textbf{22}, 7083 (2012).

\bibitem{park:2012}
H. Park, A. Afzali,	S.-J. Han, G. S. Tulevski, A. D. Franklin, J. Tersoff, J. B. Hannon, and W. Haensch,
Nature Nanotechnol. \textbf{7}, 787 (2012).

\bibitem{hermann:2010}
L. G. Hermann, F. Portier, P. Roche, A. Levy Yeyati, T. Kontos, and C. Strunk,
Phys. Rev. Lett. {\bf 104}, 026801 (2010).

\bibitem{schindele:2012}
J. Schindele, A. Baumgartner, and C. Sch\"{o}nenberger,
Phys. Rev. Lett. \textbf{109}, 157002 (2012).

\bibitem{recher:2001}
P. Recher, E. V. Sukhorukov, and D. Loss,
Phys. Rev. B \textbf{63}, 165314 (2001).

\bibitem{hofstetter:2009}
L. Hofstetter, S. Csonka, J. Nyg\aa{}rd and C. Sch\"onenberger,
Nature (London) {\bf461}, 960, (2009).

\bibitem{hofstetter:2011}
L. Hofstetter, S. Csonka, A. Baumgartner, G. F\"{u}l\"{o}p, S. d'Hollosy, 
J. Nyg{\aa}rd, and C. Sch\"{o}nenberger,
Phys. Rev. Lett. \textbf{107}, 136801 (2011).

\bibitem{das:2012}
A. Das, Y. Ronen, M. Heiblum, D. Mahalu, A. V. Kretinin, and H. Shtrikman,
Nature Comm. \textbf{3}, 1165 (2012).

\bibitem{torres:1999}
J. Torr\`{e}s and T. Martin,
Eur. Phys. J. B {\bf 12}, 319 (1999).

\bibitem{deutscher:2000}
G. Deutscher and D. Feinberg,
Appl. Phys. Lett. {\bf 76}, 487 (2000).

\bibitem{lesovik:2001}
G. Lesovik, T. Martin, and G. Blatter,
Eur. Phys. J. B {\bf 24}, 287 (2001). 

\bibitem{recher:2002}
P. Recher and D. Loss,
Phys. Rev. B {\bf 65}, 165327 (2002).

\bibitem{bena:2002}
C. Bena, S. Vishveshwara, L. Balents, and M. P. A. Fisher,
Phys. Rev. Lett. {\bf 89}, 037901 (2002).

\bibitem{recher:2003}
P. Recher, D. Loss, 
Phys. Rev. Lett. {\bf 91}, 267003 (2003).

\bibitem{hu:2004}
X. Hu and S. Das Sarma,
Phys. Rev. B \textbf{69}, 115312 (2004).

\bibitem{samuelsson:2004}
P. Samuelsson, E. V. Sukhorukov, and M. B\"{u}ttiker,
Phys. Rev. B \textbf{70}, 115330 (2004).

\bibitem{sauret:2005}
O. Sauret, T. Martin, and D. Feinberg,
Phys. Rev. B {\bf 72}, 024544 (2005).

\bibitem{bayandin:2006}
K. V. Bayandin, G. B. Lesovik, and T. Martin,
Phys. Rev. B {\bf 74}, 085326 (2006).

\bibitem{ando:2000}
T. Ando,
J. Phys. Soc. Jpn. \textbf{69}, 1757 (2000).

\bibitem{chico:2004}
L. Chico, M. P. L\'{o}pez-Sancho, and M. C. Mu\~{n}oz,
Phys. Rev. Lett. \textbf{93}, 176402 (2004).

\bibitem{huertas-hernando:2006}
D. Huertas-Hernando, F. Guinea, and A. Brataas,
Phys. Rev. B \textbf{74}, 155426 (2006).

\bibitem{izumida:2009}
W. Izumida, K. Sato, and R. Saito,
J. Phys. Soc. Jpn. \textbf{78}, 074707 (2009).

\bibitem{jeong:2009}
J.-S. Jeong and H.-W. Lee,
Phys. Rev. B \textbf{80}, 075409 (2009).

\bibitem{chico:2009}
L. Chico, M. P. L\'{o}pez-Sancho, and M. C. Mu\~{n}oz,
Phys. Rev. B {\bf 79}, 235423 (2009).

\bibitem{klinovaja:2011a}
J. Klinovaja, M. J. Schmidt, B. Braunecker, and D. Loss,
Phys. Rev. Lett. \textbf{106}, 156809 (2011).

\bibitem{klinovaja:2011b}
J. Klinovaja, M. J. Schmidt, B. Braunecker, and D. Loss,
Phys. Rev. B \textbf{84}, 085452 (2011).

\bibitem{delvalle:2011}
M. del Valle, M. Marga\'{n}ska, and M. Grifoni,
Phys. Rev. B \textbf{84}, 165427 (2011).

\bibitem{chico:2012}
L. Chico, H. Santos, M. C. Mu\~{n}oz, and M. P. L\'{o}pez-Sancho,
Solid State Commun. {\bf 152}, 1477 (2012).

\bibitem{bulaev:2008}
D. V. Bulaev, B. Trauzettel, and D. Loss,
Phys. Rev. B \textbf{77}, 235301 (2008).

\bibitem{weiss:2010}
S. Weiss, E. I. Rashba, F. Kuemmeth, H. O. H. Churchill, and K. Flensberg,
Phys. Rev. B {\bf 82}, 165427 (2010).

\bibitem{burset:2011}
P. Burset, W. Herrera, and A. Levy Yeyati, 
Phys. Rev. B {\bf 84}, 115448 (2011).

\bibitem{cottet:2012a}
A. Cottet, T. Kontos, and A. Levy Yeyati, 
Phys. Rev. Lett. {\bf 108}, 166803 (2012).

\bibitem{cottet:2012b}
A. Cottet,
Phys. Rev. B {\bf 86}, 075107 (2012).

\bibitem{braunecker:2013}
B. Braunecker, P. Burset and A. Levy Yeyati,
Phys. Rev. Lett. \textbf{111}, 136806 (2013).

\bibitem{kuemmeth:2008}
F. Kuemmeth, S. Ilani, D. C. Ralph, and P. L. McEuen,
Nature (London) \textbf{452}, 448 (2008).

\bibitem{jespersen:2011}
T. S. Jespersen, K. Grove-Rasmussen, K. Flensberg, J. Paaske, K. Muraki,
T. Fujisawa, and J. Nyg\aa{}rd,
Phys. Rev. Lett. \textbf{107}, 186802 (2011).

\bibitem{steele:2013}
G. A. Steele, F. Pei, E. A. Laird, J. M. Jol, H. B. Meerwaldt, and L. P. Kouwenhoven,
Nature Commun. {\bf 4}, 1573 (2013).

\bibitem{postma:2000}
H. W. Ch. Postma, M. de Jonge, Z. Yao, and C. Dekker,
Phys. Rev. B {\bf 62}, R10653 (2000).

\bibitem{fuhrer:2000}
M. S. Fuhrer, J. Nyg\aa{}rd, L. Shih, M. Forero, Y. G. Yoon,
M. S. C. Mazzoni, H. J. Choi, J. Ihm, S. G. Louie, A. Zettl, and P. L. McEuen,
Science {\bf 288}, 494 (2000).

\bibitem{kim:2001}
J. Kim, K. Kang, J.-O. Lee, K.-H. Yoo, J.-R. Kim, J. W. Park, 
H. M. So, and J.-J. Kim, 
J. Phys. Soc. Jpn. {\bf 70}, 1464 (2001).

\bibitem{janssen:2002}
J. W. Janssen, S. G. Lemay, L. P. Kouwenhoven, and C. Dekker,
Phys. Rev. B {\bf 65}, 115423 (2002) 

\bibitem{yoneya:2002}
N. Yoneya, K. Tsukagoshi, and Y. Aoyagi,
Appl. Phys. Lett. {\bf 81}, 2250 (2002).

\bibitem{park:2003}
J. W. Park, J. Kim, and K.-H. Yoo,
J. Appl. Phys. {\bf 93}, 4191 (2003).

\bibitem{gao:2004}
B. Gao, A. Komnik, R. Egger, D. C. Glattli, and A. Bachtold,
Phys. Rev. Lett. {\bf 92}, 216804 (2004).

\bibitem{yoon:2001}
Y.-G. Yoon, M. S. C. Mazzoni, H. J. Choi, J. Ihm, and S. G. Louie,
Phys. Rev. Lett. {\bf 86}, 688 (2001).

\bibitem{nakanishi:2001}
T. Nakanishi and T. Ando,
J. Phys. Soc. Jpn., \textbf{70}, 1647 (2001).

\bibitem{komnik:2001}
A. Komnik and R. Egger,
Eur. Phys. J. B {\bf 19}, 271 (2001).

\bibitem{dag:2004}
S. Dag, R. T. Senger, and S. Ciraci.
Phys. Rev. B {\bf 70}, 205407 (2004).

\bibitem{margulis:2007}
V. Margulis and M. Pyataev,
Phys. Rev. B {\bf 76}, 085411 (2007).

\bibitem{havu:2011}
P. Havu, M. J. Hashemi, M. Kaukonen, E. T. Sepp\"{a}l\"{a}, and R. M. Nieminen,
J. Phys.: Cond. Mat. {\bf 23}, 112203 (2011).

\bibitem{burkard:2000}
G. Burkard, D. Loss and E. V. Sukhorukov,
Phys. Rev. B {\bf 61}, 16303 (2000).

\bibitem{egues:2002}
J. C. Egues, G. Burkard, and D. Loss,
Phys. Rev. Lett. \textbf{89}, 176401 (2002).

\bibitem{egues:2005}
J. C. Egues, G. Burkard, D. S. Saraga, J. Schliemann, and D. Loss,
Phys. Rev. B \textbf{72}, 235326 (2005).

\bibitem{sanjose:2006}
P. San-Jose and E. Prada,
Phys. Rev. B \textbf{74}, 045305 (2006).

\bibitem{kawabata:2001}
S. Kawabata,
J. Phys. Soc. Jpn. \textbf{70}, 1210 (2001).

\bibitem{chtchelkatchev:2002}
N. Chtchelkatchev, G. Blatter, G. Lesovik, and T. Martin, 
Phys. Rev. B {\bf 66}, 161320 (2002).

\bibitem{samuelsson:2003}
P. Samuelsson, E. V. Sukhorukov, and M. B\"{u}ttiker,
Phys. Rev. Lett. \textbf{91}, 157002 (2003).

\bibitem{faoro:2007}
L. Faoro and F. Taddei,
Phys. Rev. B \textbf{75}, 165327 (2007).

\bibitem{samuelsson:2006}
P. Samuelsson and M. B\"{u}ttiker,
Phys. Rev. B {\bf 73}, 041305 (2006).

\bibitem{saito:1998}
R. Saito, G. Dresselhaus, and M. S. Dresselhaus,
{\it Physical Properties of Carbon Nanotubes}
(Imperial College Press, London, 1998).

\bibitem{serrano:2000}
J. Serrano, M. Cardona, and J. Ruf,
Solid State Commun. \textbf{113}, 411 (2000).

\bibitem{min:2006}
H. Min, J. E. Hill, N. A. Sinitsyn, B. R. Sahu, L. Kleinman, and A. H. MacDonald,
Phys. Rev. B \textbf{74}, 165310 (2006).

\bibitem{guinea:2010}
F. Guinea,
New J. Phys. \textbf{12}, 083063 (2010).

\bibitem{buttiker:1992}
M. B\"uttiker,
Phys. Rev. B {\bf 46}, 12485 (1992).

\bibitem{jhang:2011}
S.-H. Jhang, M. Marga\'{n}ska, M. del Valle,
Y. Skourski, M. Grifoni, J. Wosnitza, and C. Strunk,
Phys. Stat. Sol. B \textbf{248}, 2672 (2011).

\bibitem{flensberg:2010}
K. Flensberg and C. Marcus,
Phys. Rev. B \textbf{81}, 195418 (2010).

\bibitem{pei:2012}
F. Pei, E. A. Laird, G. A. Steele, and L. P. Kouwenhoven,
Nature Nanot. {\bf 7}, 630 (2012).

\bibitem{laird:2012}
E. A. Laird, F. Pei, L. P. Kouwenhoven,
Nature Nanotechnol. \textbf{8}, 565-568 (2013).

\bibitem{lai:2012}
R. A. Lai, H. O. H. Churchill, and C. M. Marcus,
arXiv:1210.6402.

\bibitem{note}
For the proposal of Ref. \onlinecite{recher:2001}, 
$\epsilon_0$ is determined by the 
quantum dot resonances. To avoid interaction with the underlying Fermi sea of the normal 
leads, which potentially could destroy the entanglement of the injected pair,
we should still have $\epsilon_0-\epsilon_F > k_B T, \Gamma_\text{inj}$, with $\Gamma_\text{inj}$ the 
tunneling rates from the quantum dots to the normal leads (\cite{recher:2001}).


\end{thebibliography}
\end{document}